\documentclass[12pt]{iopart}
\usepackage[tbtags]{amsmath}
\usepackage{amssymb}
\usepackage{physics} 
\usepackage{threeparttable}

\newcommand{\orcid}[1]{\href{https://orcid.org/#1}{\textcolor[HTML]{A6CE39}{\aiOrcid}}}

\usepackage{hyperref}% add hypertext capabilities
\usepackage{chngcntr}

\usepackage{tikz,xcolor}

\definecolor{lime}{HTML}{A6CE39}
\DeclareRobustCommand{\orcidicon}{%
	\begin{tikzpicture}
	\draw[lime, fill=lime] (0,0) 
	circle [radius=0.16] 
	node[white] {{\fontfamily{qag}\selectfont \tiny ID}};
	\draw[white, fill=white] (-0.0625,0.095) 
	circle [radius=0.007];
	\end{tikzpicture}
	\hspace{-2mm}
}

\foreach \x in {A, ..., Z}{%
	\expandafter\xdef\csname orcid\x\endcsname{\noexpand\href{https://orcid.org/\csname orcidauthor\x\endcsname}{\noexpand\orcidicon}}
}

\newcommand{\modify}[1]{\textcolor{black}{#1}}

\makeatletter
\def\@email#1#2{%
 \endgroup
 \patchcmd{\titleblock@produce}
  {\frontmatter@RRAPformat}
  {\frontmatter@RRAPformat{\produce@RRAP{*#1\href{mailto:#2}{#2}}}\frontmatter@RRAPformat}
  {}{}
}%
\makeatother

\begin{document}

\title[]{Investigation of Quasi-particle Relaxation in Strongly Disordered Superconductor Resonators}

\author{Jie Hu$^1$\orcidA{}, Jean-Marc Matin$^1$\orcidB{}, Paul Nicaise$^1$\orcidC{}, Faouzi Boussaha$^1$\orcidD{}, Christine Chaumont$^1$, Michel Piat$^2$, Pham Viet Dung$^2$ and Piercarlo Bonifacio$^3$\orcidE{}}

\address{$^1$ GEPI, Observatoire de Paris, Université PSL, CNRS, 75014 Paris, France}
\address{$^2$ Université de Paris, CNRS, Astroparticule et Cosmologie, F-75013 Paris, France}
\address{$^3$ GEPI, Observatoire de Paris, Université PSL, CNRS, 92195 Meudon, France}
\ead{jie.hu@obspm.fr}
\vspace{10pt}
\begin{indented}
\item[]December, 2023
\end{indented}

\begin{abstract}
In this paper, we investigate the quasi-particle (QP) relaxation of strongly disordered superconducting resonators under optical illumination at different bath temperatures with the Rothwarf and Taylor equations and the gap-broadening theory described by the Usadal equation. The analysis is validated with various single-photon responses of Titanium Nitride (TiN) microwave kinetic inductance detectors (MKIDs) under pulsed 405~nm laser illumination. The QP relaxation in TiN is dominated by QPs with energy below the energy gap smeared by the disorder, and its duration is still inversely proportional to the QP density. The QP lifetime versus temperature can be fitted. The relaxation of the resonator can be further modeled with QP diffusion. The fitted QP diffusion coefficient of TiN is significantly smaller than expected. Our result also shows a significant increase in QP generation efficiency as the bath temperature increases. 
\end{abstract}

%\maketitle

\section{Introduction}

Strongly disordered superconductors have been widely used for superconducting resonators for photon detection in astrophysics\cite{Leduc2010, Mazin2013, Bueno2014, Zobrist2019} and qubit readout in quantum circuits\cite{Chang2013}. They are particularly favored for microwave kinetic inductance detectors\cite{Day2003} (MKIDs) in the optical and near-infrared bands because of their high quality and high resistivity for better optical efficiency\cite{Mazin2013, Gao2012,  Guo2017}. 

Compared to aluminum superconducting resonators\cite{Visser2014_Nat}, 
MKIDs made of strongly disordered superconductors show a different QP relaxation process under optical illumination. First, the QP relaxation in these MKIDs shows a very fast QP relaxation after photon absorption as the bath temperature ($T_{bath}$) decreases\cite{Mazin2013, Gao2012,  Guo2017, Zobrist2022}. Second, their resonance frequency shift versus $T_{bath}$ is proportional to the optical power ($P_{opt}$) under steady illumination\cite{Bueno2014, Hubmayr2015}, while the resonance frequency of MKIDs made of superconductor described by BCS theory\cite{BCS1957} is proportional to\cite{Visser2014_Nat} $\sqrt{P_{opt}}$. Third, the response in the optical band is much smaller than expected\cite{Guo2017, Boussaha2023, Hu2023}.
Fourth, the density of states (DoS) of the QPs below the energy gap ($\Delta$) is not zero due to the energy gap being smeared by the disorder\cite{Driessen2012}. Thus, many QPs with energy less than $\Delta$ would be in the disordered superconductors. However, the response of these QPs to the optical illumination remains unknown. 

In this paper, we investigate the QP relaxation time in strongly disordered superconductors with the single-photon response of TiN MKIDs at 405~nm. We show that the QP relaxation in TiN under optical illumination at low temperatures still follows the Rothwarf and Taylor (RT) equations\cite{Rothwarf1967} despite the observed anomalous response. 
%The sub-gap QPs still contribute equally to the QP recombination with those QPs with energy larger than $\Delta$. 
%The QP relaxation in TiN is dominated by QPs below the energy gap smeared by the disorder. 
We have also obtained an analytical solution of RT equations for superconductors with relatively slow relaxation at low temperatures with various phonon-trapping factors. 
We further considered QP diffusion in the superconductor and fitted the diffusion coefficient ($D_{qp}$) and the pair-breaking coefficient ($\eta$), showing there is a significant increase in $\eta$ as the bath temperature increases, which can be the reason for the increasing responsivity of TiN under illumination observed elsewhere\cite{Boussaha2023}. The fitted $D_{qp}$ is significantly smaller than expected, which can be why the QPs relax quickly after the photon absorption. 
%We have further numerically solved the diffusive RT equations and show there is a significant reduction of the diffusive coefficient as the temperature goes lower, which can be the reason for the increasing responsivity of TiN under illumination observed elsewhere \cite{Boussaha2023}. 

MKIDs in the optical and near-infrared bands are usually high-quality lumped superconducting resonators\cite{Gao2012, Zobrist2022, Beldi2019, Visser2021Phonon, HuJie2021, Nicaise2022} that are made of an interdigitated capacitor (IDC) and a meander. Photons ($\hbar\omega > 2\Delta$)  absorbed in the meander break Cooper pairs and generate QPs, recombining into Cooper pairs and emitting phonons within the QP lifetime. The extra QPs increase the kinetic inductance and lower the resonance frequency. Such a response can be read out in the phase and amplitude of the resonator by a probing tone at the resonance frequency. 

\section{Superconducting Gap Broadening}
The gap broadening of disorder superconductors is formulated by the Usadal equations as\cite{Driessen2012, Coumou2013, Lyu2023}
\begin{align}\label{eqn: disorder}
    iE\sin\theta + \Delta \cos\theta - \alpha_d \sin \theta\cos\theta = 0, 
\end{align}
where $\sin\theta$ and $\cos\theta$ are the quasi-classical disorder-averaged Green's functions. $E$ is the energy relative to the Fermi level. The QP DoS can be obtained as $\rho_{qp}(E) = N_0\Re{\cos\theta}$. 
The case $\alpha_d = 0$ corresponds to the BCS DoS with a peak at $E = \Delta$. 
When $\alpha_d > 0$, the superconducting gap will be broadened, and there would be nonzero DoS for $E<\Delta$. 
It means there are QPs with $E<\Delta$. 
The complex superconductivity $\sigma = \sigma_1 - i\sigma_2$ can be obtained via Nam's theory\cite{Nam1967} as 

\begin{equation}
\begin{split}
\frac{\sigma_1}{\sigma_n}=  &\frac{1}{\hbar\omega}\int_{E_g-\hbar\omega}^{-E_g} g_1(E,E^{\prime})[1-2f(E^{\prime})]\dd E\\
 & + \int_{E_g}^\infty g_1(E,E^\prime)[f(E)-f(E^\prime)]\dd E \\
\end{split} 
\end{equation}
\begin{align}\label{eqn: Nam theory}
\begin{split}
\frac{\sigma_2}{\sigma_n} &= \frac{1}{\hbar\omega}\int_{E_g-\hbar\omega}^\infty g_2(E, E^\prime)[1-2f(E^\prime)] \dd E \\
 & + \int_{E_g}^\infty g_2(E^\prime, E)[1-2f(E)]\dd E, 
\end{split}
\end{align}
with $g_1$ and $g_2$ as 
\begin{equation*}
    \begin{split}
g_1 =& \Re{\cos\theta(E)}\Re{\cos\theta(E^\prime])} \\ 
 &+ \Re{\cos\theta(E)}\Re{\cos\theta(E^\prime)}
\end{split}
\end{equation*}
\begin{equation*}
    \begin{split}
    g_2=&\Im{\cos\theta(E)}\Re{\cos\theta(E^{\prime})} \\
    &+ \Im{i\sin\theta(E)}\Re{i\sin\theta(E^{\prime})}
    \end{split}
\end{equation*}
$E^\prime = E+ \hbar \omega$ and $f(E)$ the Fermi-Dirac distribution, and $E_g$ is the effective energy gap where the DoS starts to be positive and $\sigma_n$ is the normal-state conductivity, equations 1.3a or 1.3b in \cite{Nam1967}.

Thus, phonons with energy above $2E_g$ can break the cooper pairs in TiN but with a lower probability. 
%The fitted $\alpha_d$ is shown in Table~\ref{table: compare}.  
%and the QP lifetime is fitted with phase relaxation starting from half of the maximum of the pulse response\cite{Zobrist2022}, as shown in Fig.~\ref{fig:df_f and tau_qp}-(B).

\section{Rothwarf and Taylor Equations}
RT equations are a general phenomenological description of the dynamics between QPs and phonons for various superconductors\cite{Twerenbold1986, Lindgren1999, Demsar2003, Kabanov2005, Beck2011} as
\begin{align}\label{eqn:Rothwarf and Taylor equations}
\begin{split}
    \frac{dn}{dt} &=  I + \beta P - Rn^2 \\
    \frac{dP}{dt} &= - \frac{\beta}{2} P + \frac{Rn^2}{2} - \gamma (P-P_T^0)
\end{split}
\end{align}
where $n = n(t)$ and $P=P(t)$ are the %temporal 
concentrations of QPs and phonons in the superconductor, $\beta=2/\tau_{B}$ is the pair-breaking rate caused by the phonon absorption, \modify{$\tau_{B}$ is the time a phonon breaks a cooper pair}, and $R$ is the QP recombination rate with the creation of a phonon. 
$I$ is the background QP generation rate per unit volume.
%$D_{qp}$ and $D_{ph}$ are the diffusion coefficients for the QPs and phonons.
$P_T^0$ is the concentration of phonons in thermal equilibrium at temperature $T$ with $I=0$, $\gamma = 1/\tau_{es}$ is their decay rate \modify{$\tau_{es}$ is the time phonons takes to escape to the substrate}. $P_T^0 = Rn_T^2/\beta$ and $n_T$ is the thermally excited QP density as 
\begin{align}\label{eqn: thermal qp density}
    %n_T(T) = 2N_0\sqrt{2\pi k_BT\Delta(T)}e^{-\Delta(T)/k_BT},
    n_T(T) = 4N_0\int_{0}^\infty f(E,T)\rho(E,T)dE,
\end{align}
where $f(E,T)$ is the Fermi-Dirac distribution and $\rho(E,T)$ is the DoS of the QPs. 
%The equivalent QP temperature is defined as $n_{qp} = n_T(T_{qp})$ and 
$N_0$ is the single spin density on the Fermi level. Factor 4 accounts for two possible electronic spins and the integration of DoS from $-\infty$ to $\infty$. 
Here all the phonons with $\hbar\omega > 2E_g$ are taken into account for the broadened superconducting gap. 
And we still consider that the number of the QP generated by a photon with energy $E_{ph}$ as 
\begin{equation}
    \delta n = \frac{\eta E_{ph}}{\Delta V},
\end{equation}
with $\eta$ the pair-breaking efficiency and $V$ the volume of the superconductor, as the QP DoS shows a maximum at $E = \Delta$.

By neglecting the spatial fluctuation of the QP density in TiN\cite{Sacepe2008}, the steady-state ($dn/dt = 0$ and $dP/dt = 0$) QP density  $n_{qp}$ and $P_{T}$ can be obtained  as\cite{Rooij2020}
$\label{eqn:qpdensity}
    n_{qp} = \sqrt{\Gamma I/R + (n_T)^2},\text{and}~P_T =I/2\gamma + P_T^0,$
where $\Gamma = (1 + \beta/2\gamma)$ is the so-called phonon trapping factor\cite{Wilson2004}.

Eq.~(\ref{eqn:Rothwarf and Taylor equations}) can be approximated as\cite{Demsar2003, Kabanov2005}  
\begin{align}\label{eqn: n(t)}
    n(t) = n_1(t) + n_2(t) - n_s
\end{align}
$n_1$ describes the dynamic for the superconductor to reach a quasi-stationary state, in which the QPs density in the superconductor reaches the maximum $n_s$ and starts to recombine back to Cooper pairs.
$n_2$ describes the relatively slow procedure\cite{Kabanov2005} ($\dd^2 n/\dd t^2 \ll \dd n/\dd t$) in which the superconductor recovers from the perturbation. 
At low temperatures ($T\ll T_c$, $T_c$ is the critical temperature of the superconductor), $n_1$, $n_2$ and $n_s$ are solved as follows\cite{Kabanov2005} 
\begin{align}
    n_1(t) &= \frac{\beta}{R}\qty[-\frac{1}{4} - \frac{1}{2\xi} + \frac{1}{\xi}\frac{1}{1-Kexp(-\beta t/\xi)}] \label{eqn:n(1)}\\
    n_2(t) &= n_{qp}\qty(\frac{1 + \kappa e^{- t/\tau_{qp}}}{1 - \kappa e^{-t/\tau_{qp}}})\label{eqn:n(2)}\\
    n_s & = \frac{\beta}{4R}(\frac{2}{\xi}-1) \label{eqn: n_s}
    %\kappa &= \frac{n_s-n_{qp}}{n_s+n_{qp}} \label{eqn:params} 
    %\kappa &= \frac{n_s-n_{qp}}{n_s+n_{qp}}
\end{align} 
with the QP relaxation time $\tau_{qp}$ as
\begin{align}\label{eqn: qp lifetime}
    \tau_{qp} = \Gamma/(2n_{qp}R), 
\end{align}
and 
\begin{align}
    \frac{1}{\xi} &= \sqrt{\frac{1}{4} + \frac{2R}{\beta}\qty(n_0 + 2P_0)} \\
    K &=\frac{(4Rn_0/\beta+1) - 2\xi^{-1}}{(4Rn_0/\beta+1) + 2\xi^{-1}} 
\end{align}
%\begin{align}
%n(t) &= n_{qp}\qty( \frac{1 + \kappa e^{- t/\tau_{qp}}}{1 - \kappa e^{-t/\tau_{qp}}}), \label{eqn:n(2)}
 %\kappa &= \frac{n_s-n_{qp}}{n_s+n_{qp}} \label{eqn:kappa}
%\end{align}
$\kappa = (n_s - n_{qp})/(n_s + n_{qp})<1$ represents the magnitude of perturbation of the QP density in the superconductor. 

\begin{figure}
    \centering
    \includegraphics[width = 10cm]{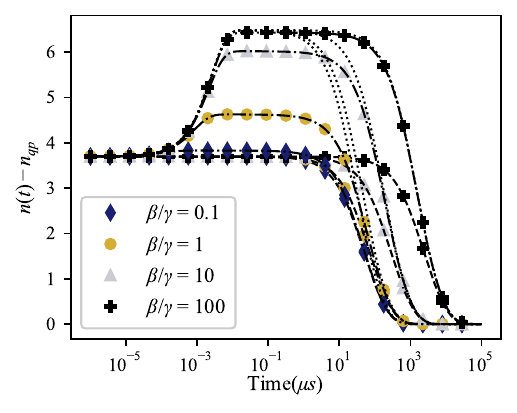}
    \caption{Analytical solution of Eq.~(\ref{eqn:Rothwarf and Taylor equations}) at low 
    %$\hdashrule[0.5ex][c]{0.4cm}{0.5pt}{1mm 2pt}$
    temperatures. ($--$), ($-\cdot$) and ($\cdot\cdot \cdot \cdot\cdot$) corresponds to $\delta P =0$, $\delta P =(1-\eta)E/2V\Delta$ and $P_0 = P_T + \delta P$ after initial photon absorption. $P_0 = P_T + \delta P$ corresponds to $\gamma/\beta=0$, which is referred as strong bottleneck condition\cite{Kabanov2005}. This condition is usually not met for superconductors at extremely low temperatures.}
    \label{fig: numerical solution of RT equation}
\end{figure}

$n_0 = n_{qp} + \delta n$ and $\delta n = \eta E_{ph}/V\Delta$ is the change of the initial QP density with $E_{ph}$ the photon energy, $\eta$ the pair-breaking efficiency and $V$ the volume of the superconductor. $P_0$ is the initial phonon density as
\begin{align}\label{eqn: phonon escape}
P_0 = P_T + \frac{\delta P}{1 + 2\gamma/\beta}
\end{align}
where $\delta P$ is the initial change of the phonon density. We introduce a factor ($ 1+2\gamma/\beta$) to account for the phonon loss when QP density reaches the maximum. 
%$\delta P$ and $\delta n = \eta E_{ph}/V\Delta$ is the initial change of the phonon density and QP density after photon absorption, where is the photon energy, $\eta$ is pair-breaking efficiency, and V is the volume of the superconductor. 
$\delta P > 0$ will lead to an increase in QP density after photon absorption\cite{Demsar2003}. 
Eq.~(\ref{eqn:n(2)}) shares the same expression by Wang et al\cite{Wang2014} (Eq.~(2)), but with a different meaning of $\kappa$. 
Eq.~(\ref{eqn:n(2)})-(\ref{eqn: phonon escape}) are valid for different phonon trapping factors and are compared with the numerical solution of Eq.~(\ref{eqn:Rothwarf and Taylor equations}), which are shown in Fig.~\ref{fig: numerical solution of RT equation}. Here the case with $\delta P=0$, $\delta P = (1-\eta)E_{ph}/(2V\Delta$) with $\eta = 0.57$ are compared, \modify{which is a typical value for Cooper-pair breaking detectors \cite{Kozorezov2000}, primarily constrained by the fact that phonons will not further break Cooper pairs when their energy is less than the binding energy of the Cooper pairs.}. 
The situation $\gamma/\beta = 0$, which corresponds to the original condition\cite{Kabanov2005} for Eq.~(\ref{eqn: n(t)}) to (\ref{eqn: qp lifetime}) is also included for comparison. This condition is usually not met for superconductors at extremely low temperatures as $\tau_{B}$ is relatively long.
The QP density increases after photon absorption for $\delta P > 0$, as observed in $\text{MgB}_2$\cite{Demsar2003}.   
Here, we set $\delta P=0$ for simplicity mainly for two reasons. First, after the superconductor absorbs a photon in the optical band, the created phonons with energy more significant than the Debye energy will escape to the substrate immediately\cite{Visser2021Phonon}. Second, the response time of MKIDs $\tau_{res} = Q/\pi f_r$ is on the order of $1~\mu \text{s}$, where $Q$ is the quality of the resonator, $f_r$ is the resonance frequency. Thus, it is difficult to determine the value of $\delta P$ from pulse relaxation. A separate measurement of a TiN film with femtosecond spectroscopy\cite{Demsar2003} is necessary to determine the value of $\delta P$, the recombination rate $R$, and the pair breaking rate $\beta$. 

\section{MKIDs Transient Response}
As the current distribution in the meander is almost uniform, the first-order time-dependent phase response of the MKID $\delta \phi$ relative to the center of its resonance circle on the IQ plane is the convolution\cite{Jonas2012, Martinez2019, kouwenhoven2022} between the detector response and the QP's temporal and spatial relaxation $n(x,t)$, which is 
\begin{align}\label{eqn:resonator_convolve}
    \delta\phi &= \int_0^t \int_V\frac{\phi_0}{\tau_{res}}e^{-(t-\tau)/\tau_{res}}\delta n(
    v, \tau) \dd v  \dd \tau, \\
    \phi_0 &= \frac{2\alpha Q \hbar \omega}{\pi\Delta_0V}\cdot \frac{\dd \sigma_2}{\dd n_{qp}}, 
\end{align}
where $\omega = 2\pi f_r$ is the angular frequency, $\alpha$ is the kinetic inductance fraction, $\Delta_0$ is the energy gap of the superconductor at absolute zero, and $\sigma_2$ is the imaginary part of the complex conductivity $\sigma = \sigma_1 - j\sigma_2$ of the superconductor. 
%$\phi_0 = \alpha S_2 Q/\Delta_0 N_0 V$ for BCS superconductor\cite{Jonas2012} and $S_2$ is the Martis-Bardeen factor. 
$\dd \sigma_2/\dd n_{qp}$ is estimated numerically with Eq.~(\ref{eqn: Nam theory}) for TiN. 
%For a BCS superconductor, $A = \alpha S_2Q/\Delta_0N_0$V, where $\alpha$ is the kinetic inductance fraction, $S_2$ is the Mattis-Bardeen factor\cite{Jonas2012} and $V$ is the volume of the meander.

\begin{table*} [h]

\begin{center}

\begin{threeparttable}

\caption{Main Parameters for TiN MKIDs}\label{table: compare}

\renewcommand\arraystretch{1.5}
%\begin{tabular}{p{2cm} p{2cm} p{1cm} p{1.3cm} p{1.3cm} p{1.3cm} p{2.3cm} p{2cm}} 
\begin{tabular}{c c c c c c c c c c c}
 \hline
    & $T_c$   & d    & $L_k$  & $\rho_n$* & $Q_i$ & $\alpha_d$ & Q & $f_r$ & $\alpha$\\ 
    &(K) & (nm) & (pH/$\square$)&($\mu \Omega \cdot $cm) & (k) & ($\Delta_0$) & (k) & (GHz) \\
 \hline\hline
$TiN_{1K}$   & 0.84    & 60      & 84   &  330 & 9.0 & 0.12  & 6.3 & 3.182 &$0.98$\\
 \hline
$TiN_{2K}$   & 2.1    & 15      & 61   &  140 & 18 & 0.092 & 13.1 & 2.625 & 0.85\\
 \hline
$TiN_{4K}$     & 4.0      & 25     & 30  & 210 & 49  &0.132 & 8.7 &4.009 & 0.60\\ 
 \hline
 
 \hline
\end{tabular}
\begin{tablenotes}
    \item \modify{*$\rho_n$ is measured at before the superconducting transition.}
\end{tablenotes}
\end{threeparttable}
\end{center}
\end{table*}

For sufficiently large $t$ (usually on the order of $10-20~\mu s$, depending on the QP lifetime) after photon absorption, the QPs diffusion can be neglected.
By inserting Eq.~(\ref{eqn: n(t)}) into Eq.~(\ref{eqn:resonator_convolve}), the temporal phase response of the resonator solved analytically is expressed as 
\begin{align}
    \delta \phi= &2n_{qp}\phi_0V\sum_{m=1}^{\infty} 
    \frac{\tau_{qp}\kappa^m}{\tau_{qp}-m\tau_{res}}\qty(e^{-mt/\tau_{qp}}-e^{-t/\tau_{res}})\label{eqn:deltaphi}
\end{align}
We neglected the contribution of $n_1(t)-n_s$ as the time scale is on the order of nanoseconds, much shorter than $\tau_{res}$. 
%$B = 2\alpha n_{qp}S_2 Q/\Delta_0N_0$. 
For $\tau_{res}\ll\tau_{qp}$, Eq.~(\ref{eqn:deltaphi}) can be further simplified as 
\begin{align}\label{eqn:qp saturation-0}
    \delta\phi \approx 2n_{qp}\phi_0V\qty(\frac{\kappa e^{-t/\tau_{qp}}}{1-\kappa e^{-t/\tau_{qp}}} - \frac{\kappa}{1-\kappa}e^{-t/\tau_{res}}).
\end{align}
The first term in Eq.~(\ref{eqn:qp saturation-0}) is the same as the result obtained by Fyhrie et al. \cite{Fyhrie2018} with an assumption that QP lifetime saturates at low temperatures\cite{Barends2009}. 
%the second term accounts for the response time of the resonator. 
Eq.~(\ref{eqn:qp saturation-0}) has been used to fit the relaxation for QP relaxation for different MKIDs\cite{Fyhrie2018, Zobrist2022}. The procedure to obtain Eq.~(\ref{eqn:resonator_convolve}) to (\ref{eqn:qp saturation-0}) can be found in \ref{sec: appendix MKIDs responsivity}

$\tau_{qp}$ fitted only by exponential decay will decrease when $\kappa$ increases. 
For $\kappa\rightarrow1$, the higher-order terms $\kappa^me^{-mt/\tau_{qp}}$ in Eq.~(\ref{eqn:deltaphi}) become dominant because they have a weight of $\kappa^m$ in the series and lead to faster relaxation of the resonator with a relaxation time of $\tau_{qp}/m$. 
Thus, the QP relaxation shows a second relaxation at low temperatures when the QP density in the superconductor is low. 
The resonator follows an exponential decay when $\kappa$ decreases, as $\kappa\rightarrow0$, only the first term in the series in Eq.~(\ref{eqn:deltaphi}) is essential. 
In this case, the QP relaxation in MKIDs approaches the exponential decay, and the relaxation time fitted from exponential decay will be close $\tau_{qp}$ as the temperature rises.

\section{MKIDs Characterization}

We investigate the QP relaxation in TiN MKIDs with different critical temperatures by fitting the single photon relaxation at $\lambda = 405$ nm. 
The detailed parameters for TiN MKIDs are listed in Table \ref{table: compare}, where $d$ is the TiN film thickness. The TiN films are deposited on sapphire substrates by magnetron sputtering, which are then patterned into resonators using photolithography and reactive ion etching. 
The films' internal quality factor $Q_i$ is relatively low \modify{as they are fabricated on sapphire\cite{Boussaha2023, Hu2023, Vissers2010}, which is probably due to the lattice mismatch on their interface}.
The MKIDs design of $TiN_{1K}$ can be found in the \ref{sec: appendix MKIDs design}
The design of $TiN_{2K}$ and $TiN_{4K}$ can be found in our previous publication\cite{Boussaha2023}. 
%Kinetic inductance $L_k$ is obtained by comparing the resonance frequency with that obtained by the electromagnetic simulation in Sonnet. 

\begin{figure}
    \centering
    \includegraphics[width = 10cm]{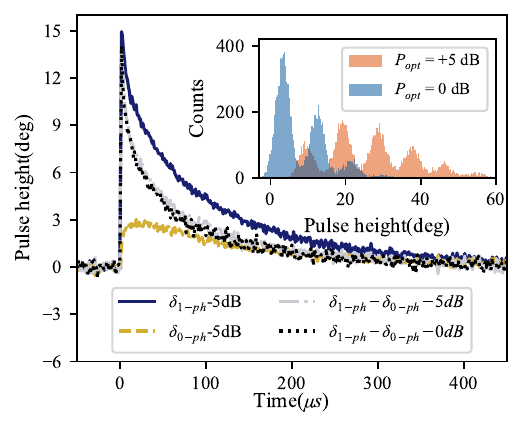}
    \caption{The 1-photon and the 0-photon responses averaged from the FWHM of the corresponding peak in the pulse statistics of the measured MKID for $P_{opt} = 5$~dBm. The photon absorption in the meander is obtained by removing the 0-photon event calculated by Eq.~(\ref{eqn:deltaphi_sum}). The inset shows the pulse statistics illuminated with two different optical powers.}
    \label{fig:0-photon removal}
\end{figure}

The MKIDs are characterized in a pulse tube-pre-cooled adiabatic demagnetization refrigerator (ADR)\cite{Hu2020}. A niobium cylinder shields the stray magnetic field and sheets of metglas 2714a around MKIDs. A pulsed 405~nm laser illuminates the MKIDs through an optical fiber. The MKIDs are read by a standard homodyne mixing scheme with a readout power of about 2~dB below bifurcation. The pulse response of the MKID is sampled by an oscilloscope at 100~MHz and processed by a Wiener optimal filter to generate the photon-counting statistics with the average pulse phase response as the template. More details on the measurement setup can be found in \ref{sec: appendix Measurement Setup}. 

The measured height pulse statistics with two different optical illumination powers are shown in the inset of Fig.~\ref{fig:0-photon removal} for $TiN_{1K}$, in which the first two peaks correspond to 0-photon and 1-photon. 
We obtain 0-photon and 1-photon responses by averaging the hundreds of events in the full-width-at-half-maximum (FWHM) in the corresponding peaks.
The 0-photon event corresponds to a non-zero background phase response that is due to electrical crosstalk from the surrounding pixels, phonons from the surrounding pixels through the substrate, as well as the photon absorbed in the Nb ground plane and the Nb feedline, as its rise time is around 3.5~$\mu \text{s}$ and is much longer than $\tau_{res}\approx 0.6~  \mu\text{s}$. The pulse statistics shift towards the right, which indicates a more significant background response. 
The 0-photon can be reduced using a microlens array\cite{Mazin2013} above the meander or continuous illumination\cite{Visser2021Phonon}. We used a pulsed illumination as it better captures the pulse response's rising edge and the multi-photon events.

Due to the high resistivity of the TiN film, the diffusion length of the QPs $l=\sqrt{D_{qp}\tau_{qp}}$ is much shorter than the total length of the meander, as the QP diffusion coefficient\cite{Narayanamurti1978} is $D_{qp} = D_N\sqrt{2k_BT/\pi\Delta}$, with $D_N = \sigma_n/e^2N_0$, $\sigma_n$ the normal state conductivity.  Therefore, the 0-photon and the photon absorption in the meander can be considered independent.  
The 1-photon response ($\delta I_1,~\delta Q_1$) can be decomposed as the summation of the 0-photon response ($\delta I_0$, $\delta Q_0$) and the photon absorption ($\delta I_{ph}$, $\delta Q_{ph}$) in the meander in the in-phase and quadrature plane as
\begin{align}
    \delta I_1 = \delta I_0 + \delta I_{ph},~~~ \delta Q = \delta Q_0 + \delta Q_{ph}.   \label{eqn:deltaphi_sum}
\end{align}
After removing $\delta I_0$ and $\delta Q_0$, the phase response of the photon absorption can be calculated, as shown in Fig. \ref{fig:0-photon removal}, which has a shorter relaxation time than the 0-photon response and a very fast relaxation after the photon absorption. And they are the same for the two optical powers.

\begin{figure}
    \centering
    \includegraphics[width = 12cm] 
    {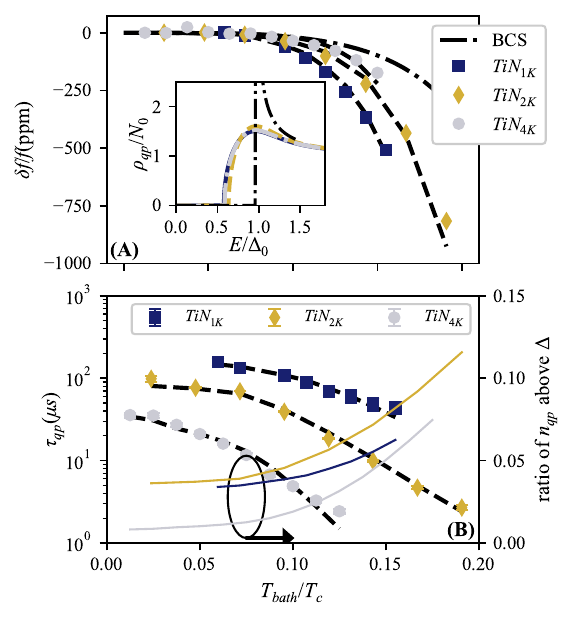}
    \caption{(A): Measured frequency shift versus $T_{bath}/T_c$ for different MKIDs as shown in Table~\ref{table: compare}. The dashed line is the fitted frequency shift with the disorder. The dashed dot line ($-\cdot$) is calculated with the standard Mattis Bardeen theory\cite{Mattis1958}. The inset shows the broadened DoS at $T = 0.1T_c$. The dashed dot line shows the DoS for the BCS superconductor. (B): Fitted QP lifetime with Eq.~(\ref{eqn:deltaphi}) from the single-photon response of different MKIDs versus $T_{bath}/T_c$. The dashed line is fitted with $\tau_{qp} = \Gamma/2n_{qp}R$ versus temperature. The right axis shows the ratio of the QPs above $\Delta$ in $n_{qp}$, calculated as $4N_0\int_{\Delta}^\infty f(E,T)\rho(E,T)dE/n_T(T_{qp})$. $n_T(T_{qp})$ is calculated with (\ref{eqn: thermal qp density}).}
    \label{fig:df_f and tau_qp}
\end{figure}

We fit the resonance frequency shift $\delta f_r/f_r$ and the QP relaxation time $\tau_{qp}$ versus $T_{bath}$ together with the superconducting gap broadening with the disorder to obtain $\Gamma/R$, $\alpha_d$, the measure of the disorder of the superconductor and a parameter to calculate the DoS of the QPs in TiN, $I$, the background QP generation rate in Eq.\~(\ref{eqn:Rothwarf and Taylor equations}) and $T_{qp}$, the temperature of the QPs. The resonance frequency shift based on the Mattis-Bardeen theory\cite{Mattis1958} is added for comparison. 
The resonance frequency versus the $T_{bath}$ can be fitted as\cite{Gao2008} $\delta f_r /f_r  = \alpha \delta \sigma_2 /2\sigma_2$ for thin film, as shown in Fig.~\ref{fig:df_f and tau_qp}-(A). The inset shows the broadened DoS at $T= T_c$. The QP lifetime shown in Fig.~\ref{fig:df_f and tau_qp}-(B) is fitted with phase relaxation starting from the half of the maximum of the pulse response\cite{Zobrist2022} with Eq.~(\ref{eqn:deltaphi}).
%The QP lifetime is fitted with Eq.~(\ref{eqn: qp lifetime}). 
Eq.~(\ref{eqn: qp lifetime}) is fitted to the measured values of the QP lifetime versus $T_{bath}$.

We assume a Kapiza boundary\cite{Pollack1969} on the interface between the TiN and the sapphire as 
\begin{align}\label{eqn: power flow}
P_d = \Sigma A_s\qty(T_{qp}^4-T_{bath}^4), 
\end{align}
where $\Sigma$ is a material constant, $A_s$ is the area of the meander and $P_d$ is the dissipated
microwave power in the resonator due to the internal loss\cite{Visser2014}.
$T_{bath}$ refers to the temperature of the sapphire substrate and is considered the same as that of the thermometer.
Here, we neglect the temperature difference between QPs and phonons in the superconductor as we do not observe significant QP lifetime saturation\cite{Guruswamy2018, Thomas2020}.
$T_{qp}$ is expected to be higher than phonon temperature in the superconductor when $T_{bath}$ is low due to the thermal decoupling between the electrons and phonons.
The obtained $T_{qp}$ is about $120$~mK, $273$~mK, and $400$~mK for $TiN_{1K}$, $TiN_{2K}$ and $TiN_{4K}$ respectively at $T_{bath} = 50$~mK. Although the phonon temperature is overestimated, the equivalent temperature of the QPs remains reasonable\cite{Budoyo2016, Guruswamy2018}. 
It can be seen that Fig.~\ref{fig:df_f and tau_qp}-(B) that the QPs with energy less than $\Delta$ dominate, and we can also expect that when $T_{bath}$ is low compared to $T_c$, the QP dynamic of TiN deviates from the BCS theory as the ratio of QPs with energy less than $\Delta$ increases. The detailed fitting procedure can be found in the \ref{sec: appendix resonance frequency shift}. 

\section{Quasi Particle Diffusion}
We further analyze the QP diffusion by introducing the diffusion term in the RT equations as\cite{Rajauria2009}
\begin{align}\label{eqn: Diffusive RT equation }
\begin{split}
    \frac{dn}{dt} &= D_{qp} \nabla^2 n + I + \beta P - Rn^2 \\
    \frac{dP}{dt} &= D_{ph} \nabla^2 P - \frac{\beta}{2} P + \frac{Rn^2}{2} - \gamma (P-P_T^0)
\end{split}
\end{align}
$D_{qp}$ and $D_{ph}$ are the diffusion coefficients for the QPs and phonons.
%The presence of the sub-gap QPs will probably reduce the responsivity of the MKIDs, as when they recombine back Cooper pairs, the phonon emitted would have energy less than $2\Delta$, which would not break the Cooper pairs further. This can be why the QP relaxation is rapid in a few microseconds after the photon absorption. It also means that when we include the sub-gap QPs in the RT equations, the phonon generation $R$ would reduce as $R$ is obtained over the QPs with different energy\cite{Chang1977}.

It is observed that as long as the QP lifetime $\tau_{qp}$ is kept the same, the solution of Eq.~(\ref{eqn:Rothwarf and Taylor equations}) will be the same for different $\Gamma$.
Thus, with the fitted $\Gamma/R$, $T_{qp}$, $I$ and $\alpha_d$, $n_{qp}$ can be calculated. $R$ can be obtained if we assume a value for $\beta$ and $\eta$. As we don't observe high energy resolving power in our MKIDs, $\Gamma$ is set to be on the order of 1. We further assume $D_{ph}=0$ the phonon diffusion is negligible. 
%With the parameters fitted from $df_r/f_r$ and $\tau_{qp}$ versus $T_{bath}$, 
In this case, by neglecting the process of 2D QP diffusion to 1D after photon absorption, the measured pulse response of MKIDs can be fitted with the result numerically obtained by solving Eq.~(\ref{eqn:Rothwarf and Taylor equations}) and (\ref{eqn:resonator_convolve}) in 1D with two fitting parameters, the quasi-particle diffusion coefficient $D_{qp}$ and the pair breaking efficiency $\eta$ with a proper initial condition.  

The initial QPs are assumed to be Gaussian distributed with $\sigma_0$ on the order of the thickness of the film. \modify{The Gaussian distribution is taken as the QP density should be continuous and symmetrical after the initial photon absorption, and it is a good approximation of the Dirac function.}
$D_{qp}$ is almost independent of $\sigma_0$, while $\eta$ shows a strong dependence as the initial relaxation rate of QP after the photon absorption is proportional to $n_{qp}^2$. %which made the value of $\eta$ meaningless. 
The $\sigma_0$ is chosen to have $\eta$ smaller than 0.6.
We consider that $\tau_{qp}$ for $TiN_{4K}$ is too short and that the initial 2D QP diffusion is essential. 
The 1D diffusion model doesn't hold for $TiN_{4K}$. 
We fix\cite{Kardakova2013} $N_0 = 6.0\times10^{10}~eV^{-1}\mu \text{m}^{-3}$, a value needed to be determined experimentally for our film by measuring the diffusion constant in the normal state.
The detailed fitting procedure can be found in \ref{sec: pulse fitting}.

\begin{figure}
    \centering
    \includegraphics[width = 10cm]{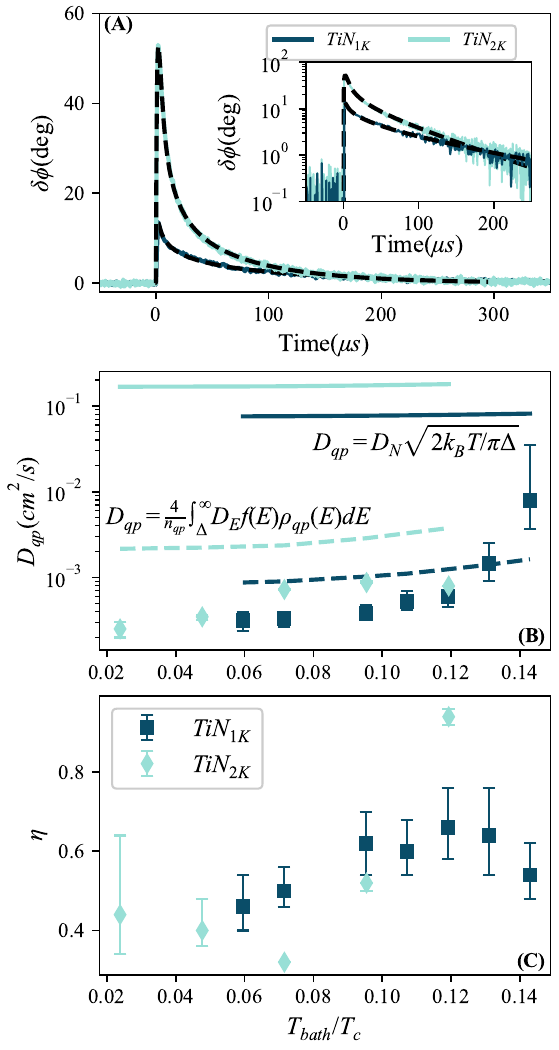}
    \caption{(A): Measured and fitted resonator relaxation for $TiN_{1K}$ and $TiN_{2K}$ at $T_{bath} = 50$~mK with 1-D diffusive RT equations. The inset shows the pulse in the log scale. (B): The fitted diffusion coefficient versus normalized temperature. The solid line is calculated as $D_{qp} = D_N\sqrt{2k_BT/\pi\Delta}$. The dashed line is calculated as $D_{qp} = \frac{4}{n_{qp}}\int_\Delta^\infty D_E f(E)\rho_{qp}(E) dE$. (C) The fitted photon to QP conversion efficiency $\eta$ at different temperatures. The error bar corresponds to one standard deviation $\sigma$ and is obtained when $\chi^2$ increases by about 2.3, corresponding to $\sigma$ for two parameter fitting.}
    \label{fig:fitted pulse response}
\end{figure}
The single photon response fitted with Eq. (\ref{eqn:Rothwarf and Taylor equations}) and (\ref{eqn:resonator_convolve}) is shown in Fig.~(\ref{fig:fitted pulse response})-(A) for $TiN_{1K}$ and $TiN_{2K}$ at $T_{bath} = 50$~mK, which shows that the fast QP relaxation term can be well modeled.

The fitted $D_{qp}$ versus $T_{bath}$ is shown in Fig.~(\ref{fig:fitted pulse response})-(B).  It is about two orders smaller than the theoretical value for two reasons. \modify{First, we have considered the QPs with energy less than $\Delta$ and still assume the QP density follows the Fermi-Dirac distribution. The number of these QPs significantly outnumbered the QPs with $E>\Delta$, as is shown in Fig.~\ref{fig:df_f and tau_qp}-(B).} The diffusion coefficient of the QPs is energy-dependent, such as for the BCS superconductor\cite{Martinis2009},  
\begin{align} \label{eqn: DE}
    D_E = D_N\sqrt{1-(\Delta/E)^2}.
\end{align}Thus, the diffusion coefficient of the low-energy QPs should be significantly lower than those with energy larger than $\Delta$. Second, there is a spatial distribution of the $\Delta$ in the TiN\cite{Escoffier2004}, which makes the QPs more difficult to diffuse. The $D_{qp}$ we obtained can be interpreted as the average $D_{qp}$ over the QPs with different energy. As QP diffusion dispersion is not known for TiN, Eq.~(\ref{eqn: DE}) is taken for calculation. Then, the averaged diffusion coefficient can be calculated as 
\begin{align}\label{eqn: D_{qp} dispersion}
    D_{qp} = \frac{4}{n_{qp}}\int_\Delta^\infty D_E f(E)\rho_{qp}(E) dE,
\end{align}
which is in qualitative agreement with the fitted QP diffusion coefficient. \modify{Essentially, we treat the QPs with $E<\Delta$ as "trapped" QPs, which are consistent with the arguments for the anomalous response in TiN\cite{Gao2012, Bueno2014}.}

A direct measurement of the QP diffusion coefficient of TiN at low temperatures is needed to investigate the diffusion coefficient further. The low diffusion coefficient can be the reason for the response of the MKIDs made of disordered superconductors being small\cite{Boussaha2023, Hu2023}. 

We focus on how $\eta$ evolves with $T_{bath}$ in Fig.~\ref{fig:fitted pulse response}-(C).    
$\eta$ tends to increase when $T_{bath}$ increases especially for $TiN_{2K}$. One of the possible reasons for the rise of $\eta$ is due to the presence of the two-level system (TLS) on the interface between the TiN and the sapphire, which absorbs part of the photon energy\cite{HuJie2021}. As $T_{bath}$ increases, TLS saturates, and then $\eta$ increases. Another possible reason is when the temperature is low, the phonon generated by the recombination of the QPs with energy less than $\Delta$ has a lower probability of breaking Cooper pairs. When $T_{bath}$ increases, the percentage of QPs with energy above $\Delta$ increases, increasing $\eta$. \modify{Additionally, an increase in $T_{bath}$ boosts the $D_{qp}$, resulting in more QPs being detected by the resonator. With the $n^2$ term influencing the recombination rate in the RT equation, a higher $D_{qp}$ translates to slower QP relaxation.} In this case, $\eta$ will increase as the temperature increases, \modify{consistent with the previous publications\cite{Bueno2014, Boussaha2023, Hu2023}, where an increase of responsivity of the MKIDs has been observed when $T_{bath}$ rises}. The fitted $\eta$ shows large uncertainty as $\eta$ is mainly related to the maximum of the pulse, which usually shows large uncertainty.

\begin{figure}
    \centering
    \includegraphics[width = 10cm]{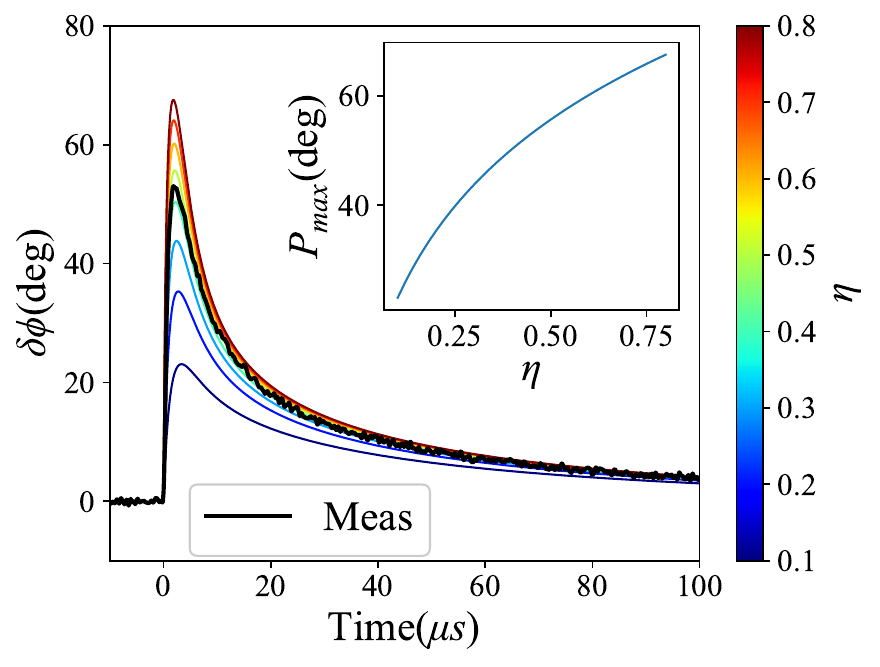}
    \caption{MKIDs response with different $\eta$ for $TiN_{2K}$ at $T_{bath} = 50$~mK. The inset shows the maximum of the pulse versus $\eta$. $D_{qp}$ is fixed with value shown in Fig.~\ref{fig:fitted pulse response}-(B)}
    \label{fig: pulse versus eta}
\end{figure}

\begin{figure}
    \centering
    \includegraphics[width =10cm]{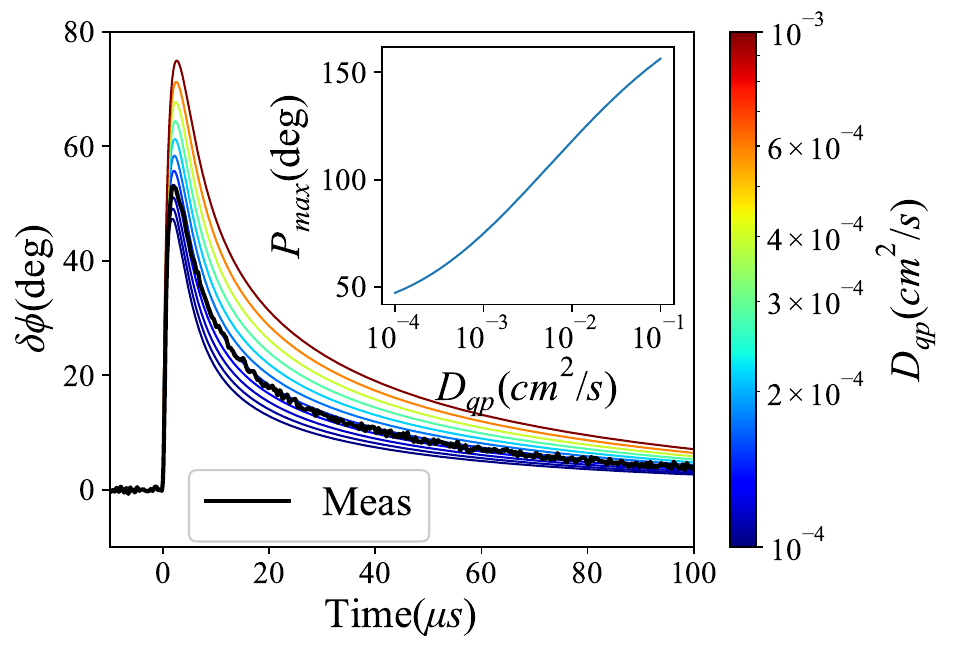}
    \caption{MKIDs response with different $D_{qp}$ for $TiN_{2K}$ at $T_{bath} = 50$~mK. The inset shows the maximum of the pulse versus $D_{qp}$. $\eta$ is fixed with value shown in Fig.~\ref{fig:fitted pulse response}-(C)}
    \label{fig: pulse versus Dqp.}
\end{figure}

We show how $\eta$ and $D_{qp}$ affect the response of MKIDs in Fig.~\ref{fig: pulse versus eta} and Fig.~\ref{fig: pulse versus Dqp.}. It can be seen that $D_{qp}$ significantly affects the pulse response.

The increase in $ \eta $ versus $T_{bath}$ could be one of the reasons why the frequency change of TiN MKIDs versus optical illumination is proportional to $P_{opt}$ in the continuous illumination in the millimeter wave range. For the continuous illumination, the change in QPs in the superconductor is $\delta n_{qp} = \eta P_{opt}\tau_{qp}/(\Delta V)$. With Eq.~(\ref{eqn: qp lifetime}), $\tau_{qp} = \Gamma / [2(n_{qp} + \delta n_{qp})R]$. Then, $\delta n_{qp}$ can be solved as 
\begin{align}
    \delta n_{qp} = -\frac{n_{qp}}{2} + \frac{1}{2}\sqrt{n_{qp}^2 + \frac{2\eta(P_{opt}) P_{opt}\Gamma}{R\Delta V}}. 
\end{align}
Here we put $\eta$ as a function of $P_{opt}$ to indicate its temperature dependence. If $\eta$ increases with $P_{opt}$, the responsivity of MKIDs would increase with $P_{opt}$, as is observed in previous publications\cite{Bueno2014, 
Hubmayr2015}.
We consider it necessary to solve the full kinetic equations\cite{Vodolazov2017} to investigate better the rising edge of the pulse response of MKIDs and better understand the temperature dependence of $\eta$.

\section{Conclusion}
In conclusion, we have modeled the single-photon relaxation process of TiN MKIDs on sapphire with the diffusive Rothwarf and Taylor equations and the gap-broadening theory based on the Usadal equation for disordered superconductors. Our results qualitatively show the QPs with energy less than $\Delta$ dominate the QP relaxation. The QP diffusion coefficient leads to fast relaxation of QPs after the photon absorption observed in different materials.  

\section*{Acknowledgement}
The authors would like to thank Damien Prêle and Manuel Gonzalez from APC, Universit{\'e} 
Paris Cit{\'e}, for discussion about measurement, as well as Florent Reix, Josiane Firminy, and Thibaut Vacelet from Paris Observatory for
assembly and mounting of the devices. The authors would also like to thank Dr. Eduard Driessen from IRAM and Dr Wei-Tao Lv from the Department of Physics at the Chinese University of Hong Kong for discussing disordered superconductors. This work is supported by the European Research Council (ERC) through Grant 835087 (SPIAKID) and the UnivEarthS Labex
program.

\appendix
\section{MKIDs Responsivity}\label{sec: appendix MKIDs responsivity}
The response of MKIDs to excess QPs can be expressed as 
\begin{align}\label{eqn: MKIDs response}
    \dd \phi = \frac{\dd \phi}{\dd\omega_r} \cdot \frac{\dd\omega_r}{\dd L} \cdot \frac{\dd L}{\dd \sigma_2} \cdot \frac{\dd \sigma_2}{\dd n_{qp}}
    \cdot \frac{\delta N_{qp}}{V}
\end{align}
with\cite{Mazin2005} $\dd \phi/ d\omega_r$ the change of phase due to the change of the resonance frequency, and $\dd \omega_r/dL$ the resonance frequency change due to the change of the inductance as 
\begin{align}
    \frac{\dd \phi}{\dd\omega_r} & = \frac{4Q}{\omega_r} \\
    \frac{\dd \omega_r}{\dd L} &= -\frac{\alpha \omega_r}{2L_k}
\end{align}
with $L = L_g +  L_k$ the total inductance in the resonator, $\alpha = L_k/L$ the fraction of the kinetic inductance and $L_g$ is the geometrical inductance. $\alpha$ is close to 1 for TiN. $\omega_r$ is the resonant frequency.  For a thin film superconductor, the kinetic inductance is $L_s = 1/(\sigma_2\omega d$) and $d$ is the film thickness\cite{Gao2008}, thus, 
\begin{align}
    \frac{\dd L}{\dd \sigma_2} & = - \frac{L_k}{\sigma_2}
\end{align}
With\cite{Gao2008} $\sigma_2/\sigma_n\approx \pi\Delta_0/(\hbar\omega)$ and $\sigma_n$ the normal state conductivity, Eq.~(\ref{eqn: MKIDs response}) can be expressed as 
\begin{align} \label{eqn: dphi appendix}
    \dd \phi  &= \frac{2\alpha Q\hbar\omega}{\pi \Delta_0\sigma_n}\cdot \frac{\dd \sigma_2}{\dd n_{qp}} \cdot \frac{\delta N_{qp}}{V} 
\end{align}
Here, $\delta N_{qp}$ is the change of the QP number in the resonator, and it can be expressed as the convolution with the detector response as
\begin{align}\label{eqn: dN_qp}
    \delta N_{qp} = \int_0^t \int_V \frac{e^{-(t-\tau)/\tau_{res}}}{\tau_{res}}\delta n(v,\tau) dx d\tau
\end{align}
where V is the volume of the meander, $\tau_{res}=Q/\pi f_r$ is the response time of the resonator. By substituting Eq.~(\ref{eqn: dphi appendix}) into Eq.~(\ref{eqn: dN_qp}), Eq.~(\ref{eqn:resonator_convolve}) in the main text can be obtained, as 
\begin{align*}
    \delta\phi &= \int_0^t \int_S\frac{\phi_0}{\tau_{res}}e^{-(t-\tau)/\tau_{res}}\delta n(
    s, \tau) \dd s  \dd \tau. \\
    \phi_0 &= \frac{2\alpha Q \hbar \omega}{\pi\Delta_0 V}\cdot \frac{\dd \sigma_2}{\dd n_{qp}}
\end{align*}
For TiN, we calculate the $\dd \sigma_2/\dd n_{qp}$ numerically with Eq.~(\ref{eqn: Nam theory}) in the main text. For a BCS superconductor, $\dd \sigma_2/ \dd n_{qp}$ can be calculated as 
\begin{align}
    \frac{\dd \sigma_2}{\dd n_{qp}} &= -\sigma_n \frac{\pi S_2}{2N_0\hbar\omega}, \\
    S_2 &= 1 + \sqrt{\frac{2\Delta_0}{\pi k_B T} }e^{-\xi}I_0(\xi)
\end{align}
with $\xi = \hbar\omega/2k_BT$ and $I_0$ is the Bessel function of the first kind. 

For sufficiently large $t$ (usually on the order of 10-20 $\mu$s,
depending on the QP lifetime) after photon absorption, the spatial distribution of the $\delta n (x, t)$ can be neglected. Thus, with Eq.~(\ref{eqn:n(2)}), $\delta n(t)$ can be expressed as 
\begin{align}\label{eqn: delta nqp uniform}
\begin{split}
    \delta n(t) &= n_{qp} \left(\frac{1 + \kappa e^{-t/\tau_{qp}}}{1-\kappa e^{-t/\tau_{qp}}} \right) - n_{qp} \\
    & = \frac{2\kappa n_{qp}e^{-t/\tau_{qp}}}{1-\kappa e^{-t/\tau_{qp}}} \\
            & = 2n_{qp}\sum_{m=1}^{\infty} \kappa^m e^{-mt/\tau_{qp}}
\end{split}
\end{align}
By substituting Eq.~(\ref{eqn: delta nqp uniform}) into Eq.~(\ref{eqn:resonator_convolve}), we obtain Eq.~(\ref{eqn:deltaphi}) as
\begin{align*}
\begin{split}
    \delta \phi &= \frac{\phi_0V}{\tau_{res}} \int_0^t e^{-(t-\tau)/\tau_{res}} 2n_{qp}\sum_{m=1}^{\infty} \kappa^m e^{-mt/\tau_{qp}} dt \\
    &= \frac{2n_{qp}V\phi_0}{\tau_{res}} \sum_{m=1}^{\infty}\kappa^m e^{-t/\tau_{res}} \int_0^t e^{-\frac{m\tau}{\tau_{qp}} + \frac{\tau}{\tau_{res}}} dt \\
     &= 2n_{qp}V\phi_0 \sum_{m=1}^{\infty} \kappa^m\frac{\tau_{qp}}{\tau_{qp}-m\tau_{res}} \left(e^{-mt/\tau_{qp}}-e^{-t/\tau_{res}} \right)
\end{split}
\end{align*}
For $\tau_{qp}\gg\tau_{res}$, we obtain Eq.~(\ref{eqn:qp saturation-0}) as 
\begin{align}
\begin{split}
    \delta \phi &\approx 2n_{qp}V\phi_0\sum_{m=1}^{\infty} \kappa^m \left(e^{-mt/\tau_{qp}}-e^{-t/\tau_{res}} \right) \\
    & = 2n_{qp}V\phi_0\qty(\frac{\kappa e^{-t/\tau_{qp}}}{1-\kappa e^{-t/\tau_{qp}}} - \frac{\kappa}{1-\kappa}e^{-t/\tau_{res}})
\end{split}
\end{align}

\section{TiN$_{1K}$ MKIDs Design}\label{sec: appendix MKIDs design}
\begin{figure}
    \centering
    \includegraphics[width = 10cm]{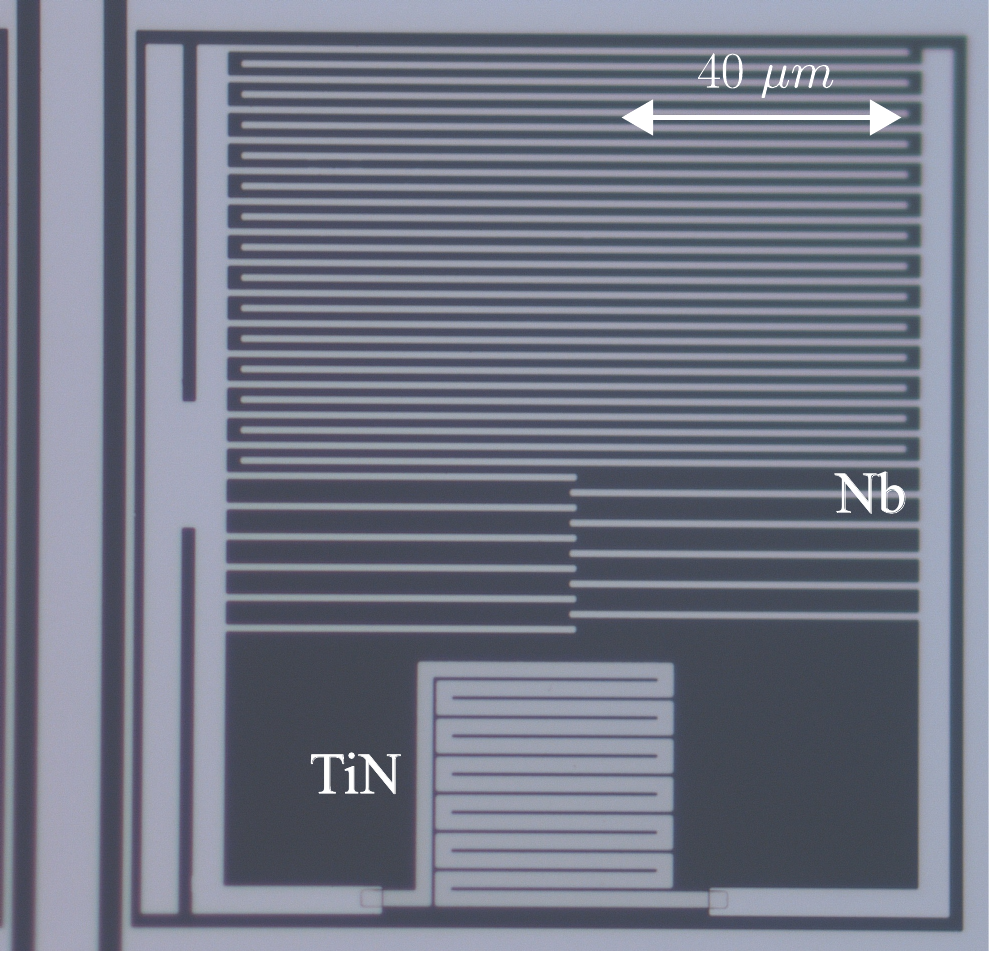}
    \caption{MKIDs design for $TiN_{1K}$. The meander is made of TiN. The rest of the resonator and the feedline are made of Niobium. }
    \label{fig: MKIDs design for 1K TiN}
\end{figure}

The detailed design of MKIDs with $TiN_{1K}$ is shown in Fig.~\ref{fig: MKIDs design for 1K TiN}. The measured TiN MKID is one of the first pixels in a 1000-pixel array. The meander line is made of sputtered 60~nm TiN on sapphire with a film resistivity of about $\rho_n = 330~ \mu \Omega \cdot \text{cm}$ ($R_s \approx 55.0~ \Omega/\square$) and its size is $40.5 \times 38.5 ~ \mu \text{m}^2$ with a strip width of $2.5 ~ \mu$m and a gap of $0.5~\mu$m. The total length of the meander is about $520~\mu m$. The IDC is made of niobium, and its size is about $120 \times 90 ~ \mu \text{m}^2$ with both strip width and spacing of $1.2 ~ \mu m$. The niobium film is 100~nm with a critical temperature of around $9.2~K$ and a kinetic inductance of about $0.2 ~pH/\square$. 

\section{Measurement Setup}\label{sec: appendix Measurement Setup}
\begin{figure*}[h]
    \centering
    \includegraphics[width = \textwidth]{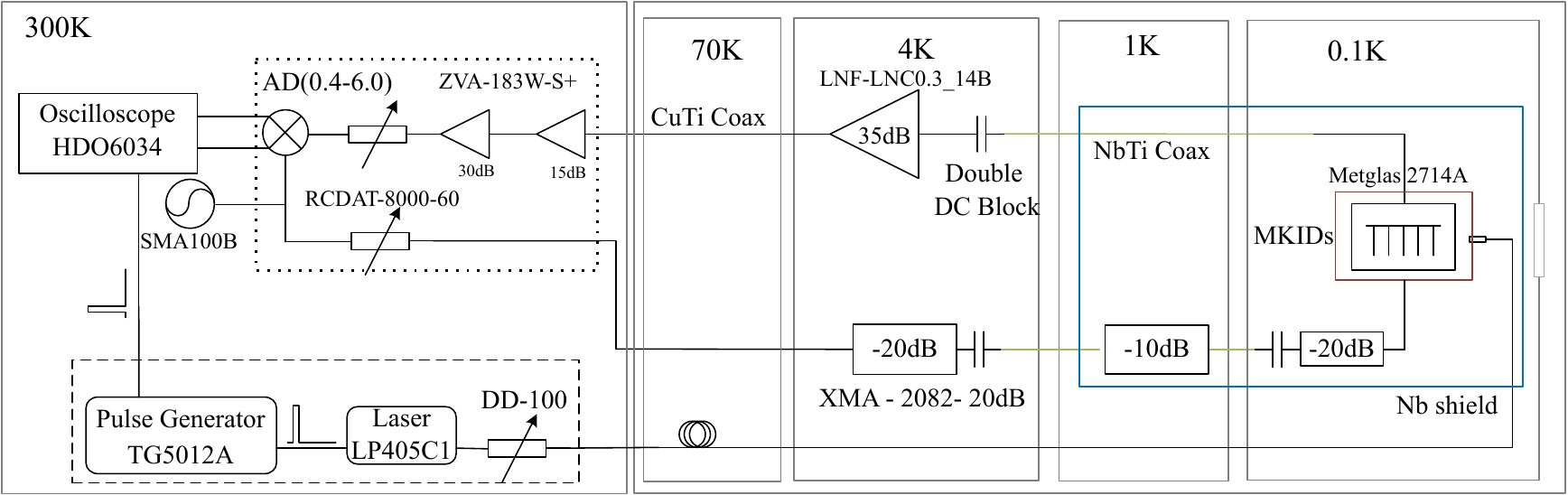}
    \caption{Detailed measurement Setup for the MKIDs}
    \label{fig:measurement setup}
\end{figure*}

The detailed measurement setup for the MKIDs is shown in Fig. \ref{fig:measurement setup}. The MKIDs are characterized in a two-stage pulse tube pre-cooled adiabatic demagnetization refrigerator (ADR) at Laboratoire Astropaticule \& Cosmologie (APC). The stray magnetic field is shielded by a niobium cylinder that is 1.5 mm thick and sheets of metglas 2714a around MKIDs. \modify{The niobium cylinder is thermalized on the 1~K stage and encloses the entire 0.1~K stage}. The MKIDs are read by a standard homodyne mixing scheme. The input signal is generated by a signal generator (SMA100A). It is first attenuated by a programmable attenuator (RCDAT-8000-60) at room temperature and then attenuated 20 dB, 10 dB, and 20 dB (XMA-2082 series) on 4 K, 1 K, and 100 mK before being fed into MKIDs. The output signal from MKIDs is first amplified by an LNA (LNF-LNC0.3\_14B) on the 4K stage and amplified further by two room temperature amplifiers (ZVA-183W-S+). The signal is down-converted to DC by an IQ mixer (AD(0.4-6.0)) and then sampled by an oscilloscope (HDO6034). Two double DC blocks (SD3463) are placed between the 4K and 1K stages to operate the heat switch in the cryostat. The readout power is estimated to be around -100 dBm at the input of the detector. The MKIDs array is illuminated by an optical fiber placed 35 mm above the pixels. The 405 nm laser (LP405C1) is modulated by a 250 Hz pulse from the pulse generator (TG5012A) of a width of 50 ns. The output power of the laser is estimated to be a few pW outside the cryostat, attenuated by a digital step attenuator (DD-100 series from OZ optics). The pulse response of the MKID is sampled by an oscilloscope (HDO6034) at 100 MHz.

\section{Resonance frequency shift and QP lifetime fitting} \label{sec: appendix resonance frequency shift}
We first fit the quality factor $Q$, resonant frequency $f_r$ and the coupling quality factor $Q_c$ from the transmission of the resonator $S_{21}$ as  
\begin{align}
S_{21} = a e^{-2\pi j f \tau}[1- \frac{Q/Q_ce^{j\phi_0}}{1+2jQ(f-f_r)/f_r}]
\end{align}
where $a$ is the amplitude of the transmission. $\tau$ is the time delay on the cable. $\phi_0$ is the parameter to account for the mismatch of the resonator to the circuit.

\begin{figure}
    \centering
    \includegraphics[width = 10cm]{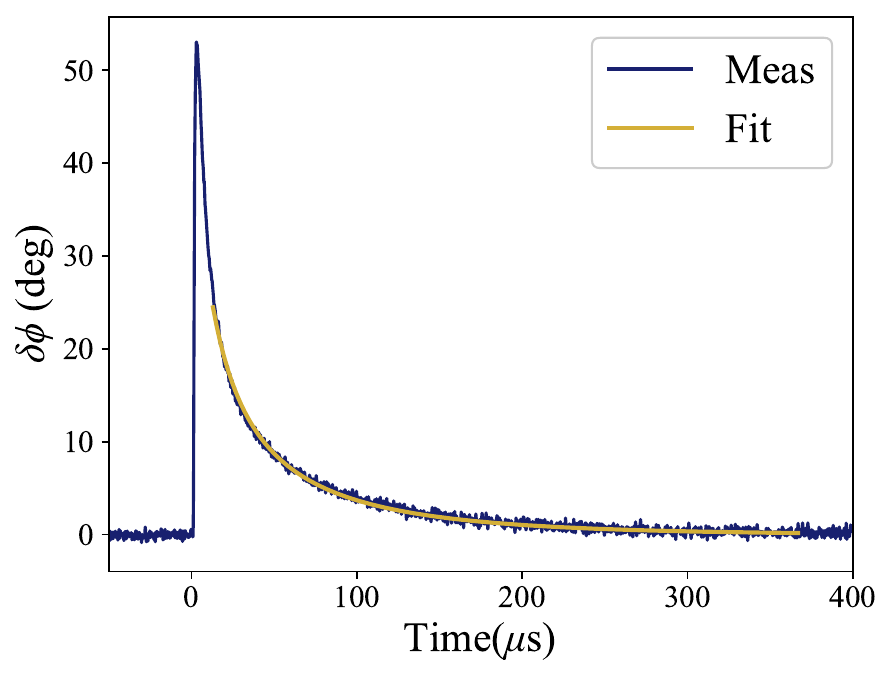}
    \caption{QP lifetime $\tau_{qp}$ fitting for $TiN_{2K}$ at $T_{bath} = 50$~mK.}
    \label{fig:Fitting QP from half maximum}
\end{figure}

The QP lifetime is fitted from the pulse relaxation from the half of the maximum using Eq.~(\ref{eqn:deltaphi}) in the main text, which we rewrite here as 
\begin{align*}
    \delta \phi &= 2Vn_{qp}\phi_0\tau_{qp}\sum_{m=1}^{\infty}\frac{\kappa^m}{\tau_{qp}-m\tau_{res}}\qty(e^{-mt/\tau_{qp}}-e^{-t/\tau_{res}}).
\end{align*}
We show an example of fitting the QP lifetime $\tau_{qp}$ for $TiN_{2K}$ at $T_{bath} = 50~$mK using Eq.~(\ref{eqn:deltaphi}) in Fig.~\ref{fig:Fitting QP from half maximum}.

We assume the QPs and the phonons to be of the same temperature and assume a Kapiza boundary on the interface between the TiN and the sapphire substrate as 
\begin{align*}
    P_{d} = \Sigma A_s\qty(T_{qp}^4-T_{bath}^4), 
\end{align*}
where $\Sigma A_s$ is a material constant, and $P$ is related to the power absorbed in the resonator\cite{Visser2014} as 
\begin{align} \label{eqn: power absorbed}
    P_{d} = \eta_r\frac{P_r}{2}\frac{4Q^2}{Q_iQ_c}\frac{Q_i}{Q_{i,qp}}
\end{align}
where $P_r$ is the readout power for the resonator, $\eta_r$ is the fraction of the energy in the resonator that is dissipated, $Q_i, Q_c, Q$ is the internal quality factor, the coupling quality factor, and the loaded quality factor respectively. $Q = Q_iQ_c/(Q_i + Q_c)$.  We consider $Q_i$ of the MKIDs to be dominated by the QPs and $Q_{i,qp} = Q_i$. 

Thus, $T_{qp}$ can be calculated as 
\begin{align}
    T_{qp} = (\frac{\eta_r}{\sum A_s}\frac{P_r}{2}\frac{4Q^2}{Q_iQ_c} + T_{bath}^4)^{1/4}
\end{align}

We set $\eta_r/\Sigma A_s$ as a fitting parameter and used $Q_i$ and $Q_c$ fitted from $S_{21}$ to obtain $T_{qp}$.

We followed the procedure in the thesis by Coumou\cite{Coumou2015} to calculate the density of states of the QPs with the disorder $\alpha_d$. We rewrite the Usadal equation Eq.~\ref{eqn: disorder} as
\begin{align*}
    iE\sin\theta + \Delta \cos\theta - \alpha_d \sin \theta\cos\theta = 0
\end{align*}
To calculate $\Delta (T)$, the Matsubara representation
of these equations is used as 
\begin{align} \label{eqn: matu represent usadal}
    \Delta \cos\theta_n - \omega_n \sin\theta_n - \alpha \sin \theta_n\cos\theta_n = 0
\end{align}
with $\theta_n = \theta(\omega_n)$ and 
\begin{align} \label{eqn: self-consistancy}
    \Delta \ln (\frac{T_c}{T}) = 2\pi k_B T \sum_{m=0}^{\infty} ( \frac{\Delta}{\omega_m} - \sin \theta_n)
\end{align}
Where $\omega_m = (2m+1)\pi k_BT$ are the Matsubara frequencies and $m = 0,~1,~2,...$. Eq.~(\ref{eqn: matu represent usadal}) and Eq.~(\ref{eqn: self-consistancy}) need to be solved iteratively for all the frequencies until the convergence is reached. Once the $\Delta(T)$ is solved, $\sin\theta(E)$ and $\cos\theta(E)$ can be obtained from Eq.~\ref{eqn: disorder}. 
With $\sin\theta(E)$ and $\cos\theta(E)$, the imagery part of complex conductivity $\sigma = \sigma_1 -j\sigma_2$ can be calculated with Eq.~(\ref{eqn: Nam theory}) as 
\begin{align*}
\begin{split}
\frac{\sigma_2}{\sigma_n} &= \int_{E_g-\hbar\omega}^\infty g_2(E, E^{\prime})[1-2f(E^{\prime})] \dd E \\
&+ \int_{E_g}^\infty g_2(E^{\prime}, E)[1-2f(E)]\dd E
\end{split}
\end{align*}
with $g_2=\Im{\cos\theta(E)}\Re{\cos\theta(E^{\prime})} + \Im{i\sin\theta(E)}\Re{i\sin\theta(E^{\prime})}$, $E=E + \hbar \omega$ and $E_g$ is effective energy gap with the DoS starts to be positive.
With the calculated density of state $\rho(E) = N_0 Re(\cos(\theta(E)))$, the QP density in the superconductor can be calculated with Eq.~(\ref{eqn: thermal qp density}) at $T_{qp}$ as 
\begin{align}
\begin{split}
    %n_T(T) = 2N_0\sqrt{2\pi k_BT\Delta(T)}e^{-\Delta(T)/k_BT},
    n_T(T_{qp}) &= 2N_0\int_{-\infty}^\infty f(E,T_{qp})\rho(E,T_{qp})dE, \\
    n_{qp} &= \sqrt{\frac{\Gamma }{R}I + n_T^2}
\end{split}
\end{align}

With Eq.~(\ref{eqn: qp lifetime}), $\tau_{qp} = \Gamma/(2n_{qp}R)$, and $\delta f/f = \alpha \delta\sigma_2/2\sigma_2$, $\tau_{qp}$ and the resonance frequency shift can be fitted together with $F_1 = \Gamma/R$, $F_2 = \eta_r/(\sum A_s)$, $F_3 = \alpha_d$ and $F_4 = I$ as the fitting parameters. We find $\gamma I/R$ is negligible.

\section{Pulse fitting}\label{sec: pulse fitting}

In this section, we describe how we fit the single photon pulse response of MKIDs. We rewrite Eq.(\ref{eqn: Diffusive RT equation }) for easy reference. 
\begin{align*}
\begin{split}
    \frac{dn}{dt} &= D_{qp} \nabla^2 n + I + \beta P - Rn^2 \\
    \frac{dP}{dt} &= D_{ph} \nabla^2 P - \beta P/2 + Rn^2/2 - \gamma (P-P_T^0)
\end{split}
\end{align*}

To solve Eq.(\ref{eqn: Diffusive RT equation }), we need to know the $D_{qp}$, $\beta$, $R$, $I$ and $P_T^0$. From the fitting of $\tau_{qp}$ versus temperature, we have obtained value $F_1 = \Gamma/R$, which is 
\begin{align}
    F_1 = (1 + \frac{\beta}{2\gamma})/R
\end{align}

\modify{
As long as we keep the same $F_1$, the solution of Eq.(\ref{eqn: Diffusive RT equation }) remains the same. This is reasonable as once $F_1$ is fixed, the $\tau_{qp}$ will be fixed with the same QP density. 
%Here, we show the simulation results with the same  $\tau_{es}/\tau_{B}$ but different $\tau_{es}$ in Fig.~. It can be seen that the results are almost the same. 
}
\modify{
We assume that $\tau_{es} = 0.35$~ns and $\tau_B = 0.25$~ns to keep the $\beta/\gamma$ on the order of 1 for both $TiN_{1K}$ and $TiN_{2K}$ as we don't observe $R$ with our detector\cite{Zobrist2022}. In Fig.~\ref{fig: comparison of tau_qp and tau_B}, we show when the $\beta/\gamma$ is kept the same, the solution is the same
\begin{figure}
    \centering
    \includegraphics{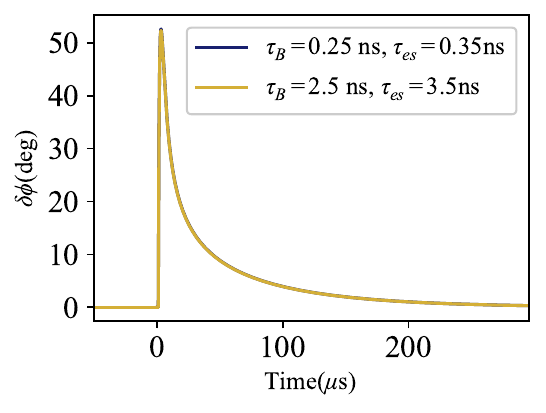}
    \caption{Comparison with simulation result with same $\beta/\gamma$ but different $\beta$ and $\gamma$}
    \label{fig: comparison of tau_qp and tau_B}
\end{figure}
$\tau_{es}$ strongly depends on the acoustic mismatch between the superconductor and the substrate\cite{Kaplan1976}. It is almost constant versus temperature\cite{Oktasendra2016}. We list the measured values $\tau_{es}$ of different commonly used superconductors for easy reference. For a film with a thickness of 60~nm, it is estimated to be 0.48~ns for niobium\cite{Oktasendra2016}, around 0.5~ns for NbN depending on substrates\cite{Sidorova2020}, and 0.19~ns for aluminum on silicon\cite{Rooij2020}. $\tau_B$ is also considered a constant as its change can be negligible in the measured temperature range\cite{Kaplan1976}. 
In this case, we obtain $\tau_0 = 248~\text{ns}$ \text{and} $ 28~\text{ns}$ for $TiN_{1K}$ and $TiN_{2K}$ respectively, to be on the order of the measured values\cite{Kardakova2013}, which is calculated as\cite{Guruswamy2018}
\begin{align}\label{eqn: R}
    \tau_0 = (\frac{2\Delta}{k_BT_c})^3\frac{1}{4\Delta N_0 R}
\end{align}
}
Here, we list the detailed parameters for fitting the pulse response in Table \ref{tab:table1}. 

With the $P_T^0 = Rn_T^2/\beta$ calculated from Eq.~(\ref{eqn: thermal qp density}), $I$ fitted \ref{sec: appendix resonance frequency shift}, and an assumption that $D_{ph} = 0$, 
we can fit Eq.(\ref{eqn: Diffusive RT equation }) for $D_{qp}$ and the pair-breaking efficiency $\eta$ with a proper initial condition. 

\begin{table*}
    \centering
    \begin{threeparttable}
    \caption{Parameters for pulse fitting for $TiN_{1K}$ and $TiN_{2K}$ @ $T_{bath} = 50~$mK}
    \begin{tabular}{c c c c}
    \hline\hline
    Parameters     &  $TiN_{1K}$  & $TiN_{2K}$ & comment \\
    \hline
    $T_c$ &0.84~K &2.1~K & The critical temperature \\ \hline
    $\alpha_d$ &0.12& 0.092& The measure of the disorder of the flim\\ \hline
    $d$ & 60~nm& 15~nm& The film thickness\\  \hline
    
    $\tau_{B}$\tnote{a} & 0.25~ns & 0.25~ns &  The pair-breaking time \\ \hline
    $f_r$ & 3.182~GHz & 2.625~GHz & The resonance frequency \\ \hline
    $\tau_{0}$  & 248~ns  & 28~ns & The electron and phonon interaction time  \\ \hline
    $\tau_{es}$\tnote{a} & 0.35~ns & 0.35~ns & The phonon escape time \\ \hline
    $Q$ &6.3k &  13.1k  &The quality factor of the resonator. \\ \hline
    $V$ &78$~\mu\text{m}^3$ & 2.55 $\mu\text{m}^3$ & The volume of the inductor. \\ \hline
    $R$ & $5.90\times10^{-18}\text{s}^{-1}$& $2.23\times10^{-17}\text{s}^{-1}$& The QP recombination rate\\ \hline
    $t_{qp}$ & 121~mK & 273~mK & The equivalent temperature of the QPs \\ \hline
    $\sigma_0$\tnote{b} &  100~nm & 60 ~nm & The variance of the initial QP distribution. \\ \hline
    \end{tabular}
    \label{tab:table1}
    \begin{tablenotes}
            \item[a] These two values are assumed to obtain the $\Gamma$. Then, by fitting the QP lifetime versus $T_{bath}$ with Eq.~(\ref{eqn: qp lifetime}), $R$ can be obtained. \modify{ $\tau_{es}$ for both films are kept the same for simplicity as we are not able to determine $\tau_{es}$ in both from our measurement.}
            \item[b] The value is obtained to set $\eta$ to be smaller than 0.6. 
        \end{tablenotes}
    \end{threeparttable}
\end{table*}

\subsection{Initial Condition}
The QPs generated by the absorbed photon in MKIDs is 
\begin{align}
    \delta N_{qp} = \frac{\eta E_{ph}}{\Delta}
\end{align}
where $E_{ph} = 3.06~\text{eV}$ for a 405~nm photon. 

We assume the QPs are Gaussian-distributed with $\sigma_0$. Then, the initial density of the QPs as 
\begin{align}
    \delta N_{qp} = wd\int_{-l/2}^{l/2} \frac{1}{\sigma_0\sqrt{2\pi}} e^{-\frac{x^2}{2\sigma_0^2}}dx
\end{align}
where $w$ is the width of the strip, $d$ is the film thickness and $l$ is the length of the meander. \modify{The Gaussian distribution is taken for simplicity as the QP density should be continuous and symmetrical after the initial photon absorption, meanwhile it is a good approximation to the Dirac function.}

%\subsection{$\tau_0$ and $\tau_{es}$ for TiN}
%To solve the RT equation, we need to know the recombination rate, $R$, the pair-breaking rate $\beta$, and the photon escape rate $\tau_{es}$. By fitting the $\tau_{qp}$ versus $T_{bath}$ and the resonance frequency shift, we obtain the $n_{qp}$ in TiN. Thus, with $\tau_{qp} = \Gamma/2n_{qp}R$. We observe that as long as we keep the ratio as $\Gamma/2n_{qp}R$, the relaxation of the resonator would be the same. In this case, we set $\tau_0 = 248~\text{ns}$ \text{and} $ 28~\text{ns}$ for $TiN_{1K}$ and $TiN_{2K}$ respectively, to be on the order of the measured values\cite{Kardakova2013}. $R$ is calculated as\cite{Guruswamy2018}
%\begin{align*}
%    R = (\frac{2\Delta}{k_BT_c})^3\frac{1}{4\Delta N_0\tau_0}
%\end{align*}

%We further assume that $\tau_{es} = 0.35$ ns for both $TiN_{1K}$ and $TiN_{2K}$, which is similar to that of aluminum. With these assumptions, we calculated the pair-breaking rate $\beta$. 

%Here, we list the detailed parameters for fitting the pulse response in Table \ref{tab:table1}.

\subsection{1D-Diffusive RT equation}
\begin{figure}
    \centering
    \includegraphics[width = 10cm]{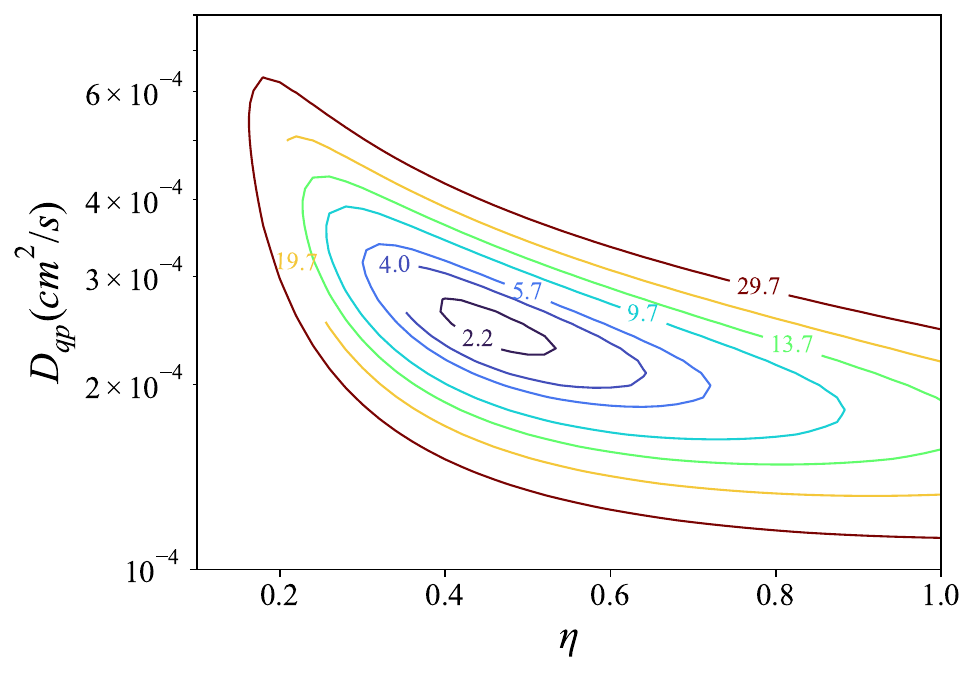}
    \caption{Fitting $\chi^2$ for $TiN_{2K}$ at $T_{bath} = 50$~mK}
    \label{fig:chi-square}
\end{figure}

We solve the 1D-Diffusive RT equation for $n_{qp}$ and keep only $\eta$ and $D_{qp}$ as the fitting parameter. We show the fitting $\chi^2$ for the pulse $TiN_{2K}$ at $T_{bath} = 50$~mK in Fig.~\ref{fig:chi-square}. 
%In Fig. \ref{fig: pulse versus eta}, we show the pulse response of MKIDs response versus $\eta$ for the same $D_{qp}$. It shows that the relationship between $\eta$ and the maximum of the pulse is not linear, which explains the large uncertainty in fitted $\eta$. In Fig. ~\ref{fig: pulse versus Dqp}, we show how the pulse evolves with different $D_{qp}$. 

\section*{References}

\bibliography{iopartnum}

\providecommand{\newblock}{}
\begin{thebibliography}{10}
\expandafter\ifx\csname url\endcsname\relax
  \def\url#1{{\tt #1}}\fi
\expandafter\ifx\csname urlprefix\endcsname\relax\def\urlprefix{URL }\fi
\providecommand{\eprint}[2][]{\url{#2}}

\bibitem{Leduc2010}
Leduc H~G, Bumble B, Day P~K, Eom B~H, Gao J, Golwala S, Mazin B~A, McHugh S, Merrill A, Moore D~C, Noroozian O, Turner A~D and Zmuidzinas J 2010 {\em Applied Physics Letters\/} {\bf 97} 102509 ISSN 0003-6951 \urlprefix\url{http://aip.scitation.org/doi/abs/10.1063/1.3480420}

\bibitem{Mazin2013}
Mazin B~A, Meeker S~R, Strader M~J, Szypryt P, Marsden D, van Eyken J~C, Duggan G~E, Walter A~B, Ulbricht G, Johnson M, Bumble B, O’Brien K and Stoughton C 2013 {\em Publications of the Astronomical Society of the Pacific\/} {\bf 125} 1348 \urlprefix\url{https://dx.doi.org/10.1086/674013}

\bibitem{Bueno2014}
Bueno J, Coumou P~C~J~J, Zheng G, de~Visser P~J, Klapwijk T~M, Driessen E~F~C, Doyle S and Baselmans J~J~A 2014 {\em Applied Physics Letters\/} {\bf 105} 192601 ISSN 0003-6951 1077-3118

\bibitem{Zobrist2019}
Zobrist N, Coiffard G, Bumble B, Swimmer N, Steiger S, Daal M, Collura G, Walter A~B, Bockstiegel C, Fruitwala N, Lipartito I and Mazin B~A 2019 {\em Applied Physics Letters\/} {\bf 115} 213503 \urlprefix\url{https://doi.org/10.1063/1.5127768}

\bibitem{Chang2013}
Chang J~B, Vissers M~R, Córcoles A~D, Sandberg M, Gao J, Abraham D~W, Chow J~M, Gambetta J~M, Beth~Rothwell M, Keefe G~A, Steffen M and Pappas D~P 2013 {\em Applied Physics Letters\/} {\bf 103} 012602 ISSN 0003-6951 \urlprefix\url{https://doi.org/10.1063/1.4813269}

\bibitem{Day2003}
Day P~K, LeDuc H~G, Mazin B~A, Vayonakis A and Zmuidzinas J 2003 {\em Nature\/} {\bf 425} 817--821 ISSN 1476-4687 \urlprefix\url{https://doi.org/10.1038/nature02037}

\bibitem{Gao2012}
Gao J, Vissers M~R, Sandberg M~O, da~Silva F~C~S, Nam S~W, Pappas D~P, Wisbey D~S, Langman E~C, Meeker S~R, Mazin B~A, Leduc H~G, Zmuidzinas J and Irwin K~D 2012 {\em Applied Physics Letters\/} {\bf 101} 142602 ISSN 0003-6951 1077-3118

\bibitem{Guo2017}
Guo W, Liu X, Wang Y, Wei Q, Wei L~F, Hubmayr J, Fowler J, Ullom J, Vale L, Vissers M~R and Gao J 2017 {\em Applied Physics Letters\/} {\bf 110} 212601 ISSN 0003-6951 1077-3118

\bibitem{Visser2014_Nat}
de~Visser P~J, Baselmans J~J, Bueno J, Llombart N and Klapwijk T~M 2014 {\em Nat Commun\/} {\bf 5} 3130 ISSN 2041-1723 (Electronic) 2041-1723 (Linking) \urlprefix\url{https://www.ncbi.nlm.nih.gov/pubmed/24496036}

\bibitem{Zobrist2022}
Zobrist N, Clay W~H, Coiffard G, Daal M, Swimmer N, Day P and Mazin B~A 2022 {\em Physical Review Letters\/} {\bf 129} 017701

\bibitem{Hubmayr2015}
Hubmayr J, Beall J, Becker D, Cho H~M, Devlin M, Dober B, Groppi C, Hilton G~C, Irwin K~D, Li D, Mauskopf P, Pappas D~P, Van~Lanen J, Vissers M~R, Wang Y, Wei L~F and Gao J 2015 {\em Applied Physics Letters\/} {\bf 106} 073505 ISSN 0003-6951 1077-3118

\bibitem{BCS1957}
Bardeen J, Cooper L~N and Schrieffer J~R 1957 {\em Physical Review\/} {\bf 108} 1175--1204 \urlprefix\url{http://link.aps.org/doi/10.1103/PhysRev.108.1175}

\bibitem{Boussaha2023}
Boussaha F, Hu J, Nicaise P, Martin J~M, Chaumont C, Dung P~V, Firminy J, Reix F, Bonifacio P, Piat M and Geoffray H 2023 {\em Applied Physics Letters\/} {\bf 122} ISSN 0003-6951 1077-3118

\bibitem{Hu2023}
Jie H, Paul N, Faouzi B, Jean-Marc M, Christine C, Alexine M, Florent R, Josiane F, Thibaut V, Dung P~V, Michel P, Elisabetta C and Piercarlo B 2024 {\em Journal of Low Temperature Physics\/} {\bf 214} 113--124 ISSN 1573-7357 \urlprefix\url{https://doi.org/10.1007/s10909-023-03018-5}

\bibitem{Driessen2012}
Driessen E~F~C, Coumou P~C~J~J, Tromp R~R, de~Visser P~J and Klapwijk T~M 2012 {\em Physical Review Letters\/} {\bf 109} 107003 \urlprefix\url{http://link.aps.org/doi/10.1103/PhysRevLett.109.107003}

\bibitem{Rothwarf1967}
Rothwarf A and Taylor B~N 1967 {\em Physical Review Letters\/} {\bf 19} 27--30 ISSN 0031-9007

\bibitem{Beldi2019}
Beldi S, Boussaha F, Hu J, Monfardini A, Traini A, Levy-Bertrand F, Chaumont C, Gonzales M, Firminy J, Reix F, Rosticher M, Mignot S, Piat M and Bonifacio P 2019 {\em Opt Express\/} {\bf 27} 13319--13328 ISSN 1094-4087 (Electronic) 1094-4087 (Linking) \urlprefix\url{https://www.ncbi.nlm.nih.gov/pubmed/31052858}

\bibitem{Visser2021Phonon}
de~Visser P~J, de~Rooij S~A, Murugesan V, Thoen D~J and Baselmans J~J 2021 {\em Phys. Rev. Applied\/} {\bf 16}(3) 034051 \urlprefix\url{https://link.aps.org/doi/10.1103/PhysRevApplied.16.034051}

\bibitem{HuJie2021}
Hu J, Boussaha F, Martin J~M, Nicaise P, Chaumont C, Beldi S, Piat M and Bonifacio P 2021 {\em Applied Physics Letters\/} {\bf 119} ISSN 0003-6951 1077-3118

\bibitem{Nicaise2022}
Nicaise P, Hu J, Martin J~M, Beldi S, Chaumont C, Bonifacio P, Piat M, Geoffray H and Boussaha F 2022 {\em Journal of Low Temperature Physics\/} {\bf 209} 1242--1248 ISSN 1573-7357 \urlprefix\url{https://doi.org/10.1007/s10909-022-02789-7}

\bibitem{Coumou2013}
Coumou P~C~J~J, Driessen E~F~C, Bueno J, Chapelier C and Klapwijk T~M 2013 {\em Physical Review B\/} {\bf 88} ISSN 1098-0121 1550-235X

\bibitem{Lyu2023}
Lyu W~T, Zhi Q, Hu J, Li J and Shi S~C 2023 {\em Chinese Physics B\/} \urlprefix\url{http://iopscience.iop.org/article/10.1088/1674-1056/ad03dc}

\bibitem{Nam1967}
Nam S~B 1967 {\em Physical Review\/} {\bf 156} 487--493 \urlprefix\url{https://link.aps.org/doi/10.1103/PhysRev.156.487}

\bibitem{Twerenbold1986}
Twerenbold D 1986 {\em Phys Rev B Condens Matter\/} {\bf 34} 7748--7759 ISSN 0163-1829 (Print) 0163-1829 (Linking) \urlprefix\url{https://www.ncbi.nlm.nih.gov/pubmed/9939456}

\bibitem{Lindgren1999}
Lindgren M, Currie M, Williams C, Hsiang T, Fauchet P, Sobolewski R, Moffat S, Hughes R, Preston J and Hegmann F 1999 {\em Applied Physics Letters\/} {\bf 74} 853--855

\bibitem{Demsar2003}
Demsar J, Averitt R~D, Taylor A~J, Kabanov V~V, Kang W~N, Kim H~J, Choi E~M and Lee S~I 2003 {\em Phys Rev Lett\/} {\bf 91} 267002 ISSN 0031-9007 (Print) 0031-9007 (Linking) \urlprefix\url{https://www.ncbi.nlm.nih.gov/pubmed/14754080}

\bibitem{Kabanov2005}
Kabanov V~V, Demsar J and Mihailovic D 2005 {\em Phys Rev Lett\/} {\bf 95} 147002 ISSN 0031-9007 (Print) 0031-9007 (Linking) \urlprefix\url{https://www.ncbi.nlm.nih.gov/pubmed/16241687}

\bibitem{Beck2011}
Beck M, Klammer M, Lang S, Leiderer P, Kabanov V~V, Gol’Tsman G and Demsar J 2011 {\em Physical Review Letters\/} {\bf 107} 177007

\bibitem{Sacepe2008}
Sacepe B, Chapelier C, Baturina T~I, Vinokur V~M, Baklanov M~R and Sanquer M 2008 {\em Phys Rev Lett\/} {\bf 101} 157006 ISSN 0031-9007 (Print) 0031-9007 (Linking) \urlprefix\url{https://www.ncbi.nlm.nih.gov/pubmed/18999631}

\bibitem{Rooij2020}
de~Rooij S 2020 {\em Quasiparticle Dynamics in Optical MKIDs: Single Photon Response and Temperature Dependent Generation-Recombination Noise\/} Master thesis TU Delft

\bibitem{Wilson2004}
Wilson C~M and Prober D~E 2004 {\em Physical Review B\/} {\bf 69} ISSN 1098-0121 1550-235X

\bibitem{Wang2014}
Wang C, Gao Y~Y, Pop I~M, Vool U, Axline C, Brecht T, Heeres R~W, Frunzio L, Devoret M~H, Catelani G, Glazman L~I and Schoelkopf R~J 2014 {\em Nat Commun\/} {\bf 5} 5836 ISSN 2041-1723 (Electronic) 2041-1723 (Linking) \urlprefix\url{https://www.ncbi.nlm.nih.gov/pubmed/25518969}

\bibitem{Kozorezov2000}
Kozorezov A~G, Volkov A~F, Wigmore J~K, Peacock A, Poelaert A and den Hartog R 2000 {\em Physical Review B\/} {\bf 61} 11807--11819 pRB \urlprefix\url{https://link.aps.org/doi/10.1103/PhysRevB.61.11807}

\bibitem{Jonas2012}
Zmuidzinas J 2012 {\em Annual Review of Condensed Matter Physics\/} {\bf 3} 169--214 ISSN 1947-5454 \urlprefix\url{http://dx.doi.org/10.1146/annurev-conmatphys-020911-125022}

\bibitem{Martinez2019}
Martinez M, Cardani L, Casali N, Cruciani A, Pettinari G and Vignati M 2019 {\em Physical Review Applied\/} {\bf 11} ISSN 2331-7019

\bibitem{kouwenhoven2022}
Kouwenhoven K, Fan D, Biancalani E, de~Rooij S~A, Karim T, Smith C~S, Murugesan V, Thoen D~J, Baselmans J~J and de~Visser P~J 2022 {\em arXiv preprint arXiv:2207.05534\/} \urlprefix\url{https://arxiv.org/abs/2207.05534}

\bibitem{Fyhrie2018}
Fyhrie A, Zmuidzinas J, Glenn J, Day P, LeDuc H~G and McKenney C 2018 Progress towards ultra sensitive kids for future far-infrared missions: a focus on recombination times vol 10708 ed Zmuidzinas J and Gao J~R International Society for Optics and Photonics (SPIE) p 107083A \urlprefix\url{https://doi.org/10.1117/12.2312867}

\bibitem{Barends2009}
Barends R, Baselmans J~J, Yates S~J, Gao J~R, Hovenier J~N and Klapwijk T~M 2008 {\em Phys Rev Lett\/} {\bf 100} 257002 ISSN 0031-9007 (Print) 0031-9007 (Linking) \urlprefix\url{https://www.ncbi.nlm.nih.gov/pubmed/18643694}

\bibitem{Vissers2010}
Vissers M~R, Gao J, Wisbey D~S, Hite D~A, Tsuei C~C, Corcoles A~D, Steffen M and Pappas D~P 2010 {\em Applied Physics Letters\/} {\bf 97} 232509 ISSN 0003-6951 1077-3118

\bibitem{Hu2020}
Hu J, Salatino M, Traini A, Chaumont C, Boussaha F, Goupil C and Piat M 2020 {\em Journal of Low Temperature Physics\/} {\bf 199} 355--361 ISSN 1573-7357 \urlprefix\url{https://doi.org/10.1007/s10909-019-02313-4}

\bibitem{Narayanamurti1978}
Narayanamurti V, Dynes R~C, Hu P, Smith H and Brinkman W~F 1978 {\em Physical Review B\/} {\bf 18} 6041--6052 \urlprefix\url{https://link.aps.org/doi/10.1103/PhysRevB.18.6041}

\bibitem{Mattis1958}
Mattis D~C and Bardeen J 1958 {\em Physical Review\/} {\bf 111} 412--417 \urlprefix\url{http://link.aps.org/doi/10.1103/PhysRev.111.412}

\bibitem{Gao2008}
Gao J, Zmuidzinas J, Vayonakis A, Day P, Mazin B and Leduc H 2008 {\em Journal of Low Temperature Physics\/} {\bf 151} 557--563 ISSN 0022-2291 1573-7357

\bibitem{Pollack1969}
Pollack G~L 1969 {\em Reviews of Modern Physics\/} {\bf 41} 48--81 rMP \urlprefix\url{https://link.aps.org/doi/10.1103/RevModPhys.41.48}

\bibitem{Visser2014}
de~Visser P~J, Goldie D~J, Diener P, Withington S, Baselmans J~J and Klapwijk T~M 2014 {\em Phys Rev Lett\/} {\bf 112} 047004 ISSN 1079-7114 (Electronic) 0031-9007 (Linking) \urlprefix\url{https://www.ncbi.nlm.nih.gov/pubmed/24580483}

\bibitem{Guruswamy2018}
Guruswamy T 2018 {\em Nonequilibrium behaviour and quasiparticle heating in thin film superconducting microwave resonators\/} Ph.d. thesis University of Cambridge

\bibitem{Thomas2020}
Thomas C~N, Withington S, Sun Z, Skyrme T and Goldie D~J 2020 {\em New Journal of Physics\/} {\bf 22} 073028 ISSN 1367-2630

\bibitem{Budoyo2016}
Budoyo R~P, Hertzberg J~B, Ballard C~J, Voigt K~D, Kim Z, Anderson J~R, Lobb C~J and Wellstood F~C 2016 {\em Physical Review B\/} {\bf 93} ISSN 2469-9950 2469-9969

\bibitem{Rajauria2009}
Rajauria S, Courtois H and Pannetier B 2009 {\em Physical Review B\/} {\bf 80} ISSN 1098-0121 1550-235X

\bibitem{Kardakova2013}
Kardakova A, Finkel M, Morozov D, Kovalyuk V, An P, Dunscombe C, Tarkhov M, Mauskopf P, Klapwijk T~M and Goltsman G 2013 {\em Applied Physics Letters\/} {\bf 103} ISSN 0003-6951 1077-3118

\bibitem{Martinis2009}
Martinis J~M, Ansmann M and Aumentado J 2009 {\em Phys Rev Lett\/} {\bf 103} 097002 ISSN 0031-9007 (Print) 0031-9007 (Linking) \urlprefix\url{https://www.ncbi.nlm.nih.gov/pubmed/19792820}

\bibitem{Escoffier2004}
Escoffier W, Chapelier C, Hadacek N and Vill\'egier J~C 2004 {\em Phys. Rev. Lett.\/} {\bf 93}(21) 217005 \urlprefix\url{https://link.aps.org/doi/10.1103/PhysRevLett.93.217005}

\bibitem{Vodolazov2017}
Vodolazov D~Y 2017 {\em Physical Review Applied\/} {\bf 7} ISSN 2331-7019

\bibitem{Mazin2005}
Mazin B~A 2005 {\em Microwave kinetic inductance detectors\/} Ph. d thesis California Institute of Technology

\bibitem{Coumou2015}
Coumou P~C~J~J 2015 {\em Electrodynamics of strongly disordered superconductors\/} Ph. d thesis TU Delft, Delft University of Technology

\bibitem{Kaplan1976}
Kaplan S~B, Chi C~C, Langenberg D~N, Chang J~J, Jafarey S and Scalapino D~J 1976 {\em Physical Review B\/} {\bf 14} 4854--4873 \urlprefix\url{https://link.aps.org/doi/10.1103/PhysRevB.14.4854}

\bibitem{Oktasendra2016}
Oktasendra F, Berdiyorov G, Mekki A and Maneval J 2016 {\em IEEE Transactions on Applied Superconductivity\/} {\bf 26} 1--4 ISSN 1051-8223

\bibitem{Sidorova2020}
Sidorova M, Semenov A, Hübers H~W, Ilin K, Siegel M, Charaev I, Moshkova M, Kaurova N, Goltsman G~N and Zhang X 2020 {\em Physical Review B\/} {\bf 102} 054501

\end{thebibliography}

\end{document}